\documentclass[manuscript]{aastex62}
\usepackage{textcomp}

\newcommand{\oq}{\textquotedblleft}
\newcommand{\cq}{\textquotedblright}
\newcommand{\cqb}{{\textquotedblright}~}    
\newcommand{\ms}{M$_\odot$}     
\newcommand{\msb}{M$_\odot$~}
\newcommand{\al}{$^{26}$Al}
\newcommand{\alb}{$^{26}$Al~}
\newcommand{\fe}{$^{60}$Fe}
\newcommand{\feb}{$^{60}$Fe~}
\newcommand{\cs}{$^{135}$Cs}
\newcommand{\csb}{$^{135}$Cs~}

\newcommand{\ca}{$^{41}$Ca}
\newcommand{\cab}{$^{41}$Ca~}

\newcommand{\mn}{$^{53}$Mn}
\newcommand{\pd}{$^{107}$Pd}
\newcommand{\pdb}{$^{107}$Pd~}

\newcommand{\hf}{$^{182}$Hf}
\newcommand{\pb}{$^{205}$Pb}
\newcommand{\ct}{$^{13}$C}

\newcommand{\cd}{$^{12}$C}

\newcommand{\ctb}{$^{13}$C~}
\newcommand{\ctanb}{$^{13}$C($\alpha$,n)$^{16}$O~}
\newcommand{\ctan}{$^{13}$C($\alpha$,n)$^{16}$O}
\newcommand{\neanb}{$^{22}$Ne($\alpha$,n)$^{25}$Mg~}

\begin{document}

\title{\Large {On the Origin of the Early Solar System Radioactivities.\\
Problems with the AGB and Massive Star Scenarios}}

\author{D. Vescovi}
\affil{Gran Sasso Science Institute, Viale Francesco Crispi, 7, 67100 L'Aquila, Italy}
\affiliation{INFN, Section of Perugia, Via A. Pascoli snc, 06123 Perugia, Italy}

\author{M. Busso}
\affiliation{INFN, Section of Perugia, Via A. Pascoli snc, 06123 Perugia, Italy}
\affiliation{University of Perugia, Department of Physics and Geology, Via A. Pascoli snc, 06123 Perugia, Italy}

\author{S. Palmerini}
\affiliation{INFN, Section of Perugia, Via A. Pascoli snc, 06123 Perugia, Italy}
\affiliation{University of Perugia, Department of Physics and Geology, Via A. Pascoli snc, 06123 Perugia, Italy}

\author{O. Trippella}
\affiliation{INFN, Section of Perugia, Via A. Pascoli snc, 06123 Perugia, Italy}

\author{S. Cristallo}
\affiliation{INFN, Section of Perugia, Via A. Pascoli snc, 06123 Perugia, Italy}
\affiliation{INAF, Observatory of Abruzzo, Via Mentore Maggini snc, 64100 Collurania, Teramo, Italy}

\author{L. Piersanti}
\affiliation{INFN, Section of Perugia, Via A. Pascoli snc, 06123 Perugia, Italy}
\affiliation{INAF, Observatory of Abruzzo, Via Mentore Maggini snc, 64100 Collurania, Teramo, Italy}

\author{A. Chieffi}
\affiliation{INAF - Istituto di Astrofisica e Planetologia Spaziali, Via Fosso del Cavaliere 100, I-00133, Roma, Italy}
\affiliation{Monash Centre for Astrophysics (MoCA),
School of Mathematical Sciences, Monash University, Victoria 3800, Australia}

\author{M. Limongi}
\affiliation{INAF - Osservatorio Astronomico di Roma, Via Frascati 33, I-00040, Monteporzio Catone, Italy}
\affiliation{Kavli Institute for the Physics and Mathematics of the Universe, Todai Institutes for Advanced Study, the University of Tokyo, Kashiwa, Japan 277-8583 (Kavli IPMU, WPI)}

\author{P.Hoppe}
\affiliation{Max-Planck-Institut f\"ur Chemie, Hahn-Meitner-Weg 1, 55128 Mainz, Germany}

\author{K.-L. Kratz}
\affiliation{Fachbereich Chemie, Pharmazie \& Geowissenschaften, Mainz Universit\"at, Fritz-Strassmann-Weg 2, 55128 Mainz, Germany}

\correspondingauthor{D. Vescovi}
\email{diego.vescovi@gssi.it}

\begin{abstract}
Recent improvements in stellar models for intermediate-mass and massive stars are recalled, together with their expectations for the synthesis of radioactive nuclei of lifetime $\tau \lesssim 25$ Myr, in order to re-examine the origins of now extinct radioactivities, which were alive in the solar nebula. The Galactic inheritance broadly explains most of them, especially if $r$-process nuclei are produced by neutron star merging according to  recent models. Instead, \al, \cab, \csb and possibly \feb require nucleosynthesis events close to the solar formation.  We outline the persisting difficulties to account for these nuclei by Intermediate Mass Stars (2 $\lesssim $ M/M$_\odot \lesssim 7 - 8$). Models of their final stages now predict the ubiquitous formation of a $^{13}$C reservoir as a neutron capture source; hence, even in presence of $^{26}$Al production from Deep Mixing or Hot Bottom Burning, the ratio $^{26}$Al/$^{107}$Pd remains incompatible with measured data, with a large excess in $^{107}$Pd. This is shown for two recent approaches to Deep Mixing. Even a late contamination by a Massive Star meets problems. In fact, inhomogeneous addition of Supernova debris predicts non-measured excesses on stable isotopes. Revisions invoking specific low-mass supernovae and/or the sequential contamination of the pre-solar molecular cloud might be affected by similar problems, although our conclusions here are weakened by our schematic approach to the addition of SN ejecta. The limited parameter space remaining to be explored for solving this puzzle is discussed.
\end{abstract}


\section{Short-lived Radioactivities in the ESS}
\label{sec:intro}
Measurements revealing that several radioactive
species with half-lives ranging from less than one to hundreds of million years were present alive in solids of the Early Solar System (ESS)
gradually accumulated in the past decades, after the pioneering work by \citet{rey} on $^{129}$I. These nuclei, in the present context, are referred to as {\it short-lived radioactivities} (SLRs).

\begin{figure*}[t!]  
\begin{center}
\includegraphics[width=\textwidth]{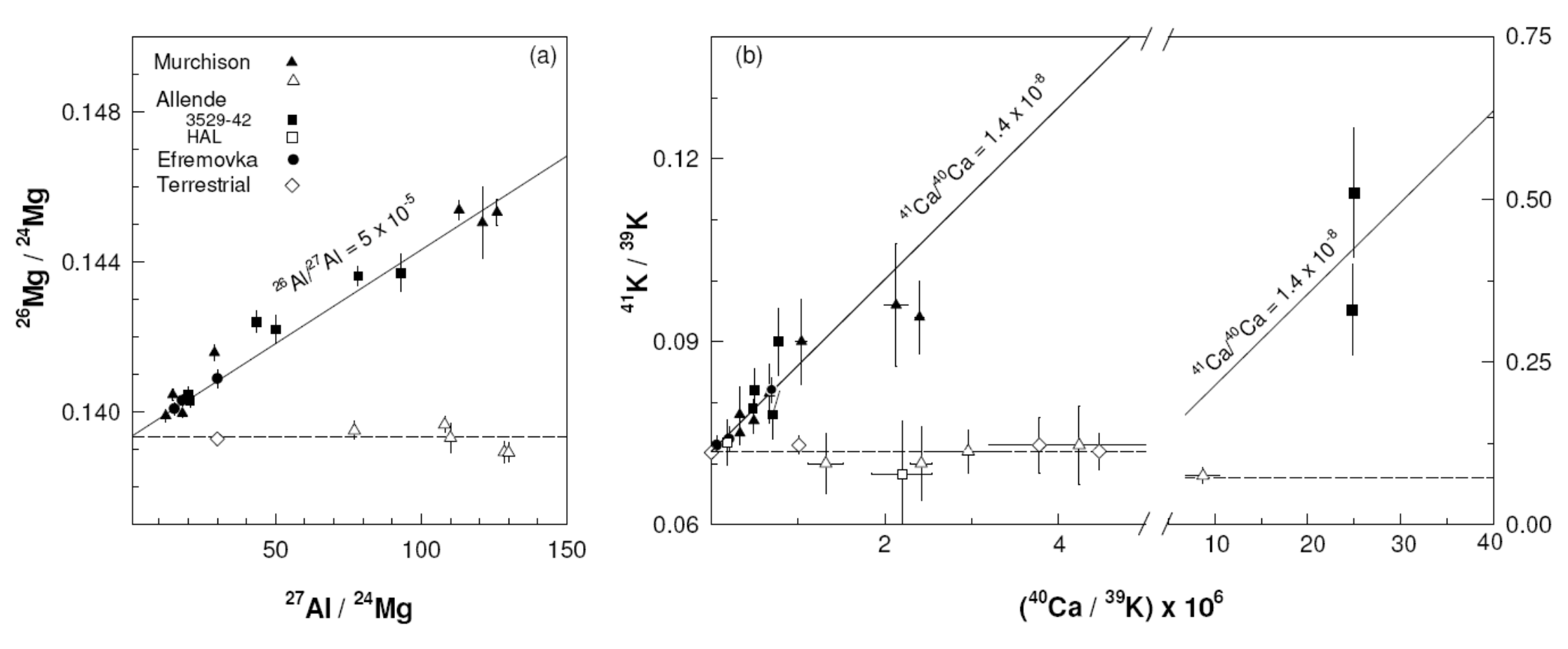}
\vskip 0pt 
\caption{Panel a) shows the Al-Mg data from Calcium and Aluminum Inclusions (CAIs) in various meteorites; the straight-line fit to the filled symbols is for the so-called canonical value of ($^{26}$Al/$^{27}$Al)$_0$ in the ESS, $\simeq 5\cdot10^{-5}$. Panel b) shows the Ca-K data from the same samples used for Panel a). First measurements and the line pointing to ($^{41}$Ca/$^{40}$Ca)$_0$ = 1.4$\cdot$10$^{-8}$ were due to \citet{sug94} and \citet{sri96}. The suggestion of a correlation with Al was due to \citet{s+98}. This figure is reproduced from these last authors and from \citet{w+06}. Copyright: Nature Publishing Group. \citep[Note that subsequent work by][proved that the initial \cab abundance was lower, at the level of ($^{41}$Ca/$^{40}$Ca)$_0  \backsim 4.2 \cdot 10^{-9}$]{liu12}.}
\label{fig:one}
\end{center}
\end{figure*}

Identification today of the stable decay product (and of its abundance) for a nucleus of this kind, permits then to extrapolate backwards,  deriving the original abundance of the unstable isotope \citep{b+99,w+06,dm14}. A description of the procedure can be found, e.g., in ~\citet{lee}. Figure \ref{fig:one} shows this technique, as applied to $^{26}$Al and $^{41}$Ca; the figure is taken from the work by \citet{s+98}. 

The wealth of new measurements on ESS samples has now become impressive. It soon posed the crucial question of which was the astrophysical interpretation of their presence: this problem is still partly unsolved now. 
It is in any case evident that, although solid materials formed in a very short lapse of time (from a fraction of a Myr to a few Myrs), they somehow maintain the records of several phenomena, from the possible blend of various stellar nucleosynthesis processes \citep{b11,b18}, to traces of spallation effects induced in the solar nebula itself \citep{sossi}. For a detailed account of the science built in the last 60 years with short-lived nuclei we refer to dedicated reviews -- see e.g. \citet{b+99,kratz,w+06, wd07,hus07,d+14,dm14}. 

Some of the experimental estimates for early SLR abundances are still uncertain and subject to  discussions. This often occurs for nuclei of metals that are extremely difficult or impossible to measure in pristine refractory condensates, so that their inferred initial abundances in the ESS require extrapolations from measurements made on subsequent, differentiated objects. In this field it is of particular importance the case of $^{60}$Fe. The values derived for the initial abundance ratio $^{60}$Fe/$^{56}$Fe wandered across the large range from a few 10$^{-9}$ \citep{td12} to about $10^{-6}$ \citep{mos05}. The review by \citet{dm14} discusses in some detail how the inference of an initial ratio, based on absolute age determinations for the eucrite samples where it was early measured, are bound to suffer for large systematic uncertainties. Those authors privileged the lowest values presented in the literature; however, there does not seem to exist a general agreement on this, especially in view of the possible large internal modifications of the abundances in the samples, subsequent to the Solar System formation \citep{tel16}; see also \citet{boss17}. The most recent measurements by \citet{fe60} stay in the lower part of the spread, but suggest an initial abundance considerably higher than the choice by \citet{dm14}.  We shall therefore consider this isotopic ratio as being uncertain inside the wide range 10$^{-8} - 10^{-6}$. 

In Table 1 we try to fix reasonable ESS abundance ratios ([$N^R/N^S$]$_{Meas.}$) for SLRs with mean life ($\tau_R$) lower than 25 Myr. We also include $^{146}$Sm as an example of longer lived nuclei. We adopt mainly \citet{dm14} as a reference; this will be in general adequate for our purposes, with the mentioned exception of $^{60}$Fe.

\begin{table*}[t!!]
\begin{center}
{
{\textsc Table 1. Short-Lived Nuclei in the ESS (*)}

\vspace{0.5cm}

\begin{tabular}{c c c c}
\hline
\hline
  Rad. & Ref. &$\tau_R$ (Myr) & $[N^R/N^S]_{\rm Meas.}$   \\
\hline
  $^{10}$Be & $^{9}$Be  & 2.0 &  ($8.8 \pm 0.6$)$\cdot10^{-4}$  \\
  $^{26}$Al & $^{27}$Al  & 1.03 & ($5.23 \pm 0.13$)$\cdot10^{-5}$  \\
  $^{36}$Cl & $^{35}$Cl  & 0.43 &  1.8$\cdot 10^{-5}$   \\
  $^{41}$Ca & $^{40}$Ca  & 0.15  & $4\cdot10^{-9}$   \\
  $^{53}$Mn & $^{55}$Mn  & 5.3 & ($6.7 \pm 0.56$)$\cdot10^{-6}$   \\
  $^{60}$Fe & $^{56}$Fe  & 3.75 & $10^{-8} - 10^{-6}$ \\
  $^{107}$Pd & $^{108}$Pd & 9.4 & ($5.9 \pm 2.2$)$\cdot 10^{-5}$   \\
  $^{129}$I & $^{127}$I  & 23 & $1.0\cdot10^{-4}$   \\
  $^{135}$Cs & $^{133}$Cs  & 3.3 & $ 4.8\cdot10^{-4}$\\
  $^{182}$Hf & $^{180}$Hf  & 12.8 & ($9.81 \pm 0.41$)$\cdot 10^{-5}$ \\
  $^{205}$Pb &  $^{204}$Pb & 22 & $ 10^{-3}$\\
  $^{247}$Cm & $^{232}$Th  & 23 & ($1.1 - 2.4$)$\cdot10^{-3}$\\
\hline
  $^{146}$Sm & $^{144}$Sm & 148 & $1.0\cdot10^{-2}$ \\

\hline
\end{tabular}
}
\end{center}
{(*) For general references see \citet{w+06} and \citet{dm14}. Among recently revised abundances, for $^{10}$Be see \citet{ch06}; for $^{60}$Fe see \citet{td12}; for $^{107}$Pd see \citet{schon08}; for $^{135}$Cs see \citet{hy13}; for $^{182}$Hf see \citet{bur08}; for $^{205}$Pb see \citet{b+10}; for $^{247}$Cm, see \citet{br10}.\\}

\label{tab1}
\end{table*}

In the last 40 years a wealth of astrophysical models were presented in the attempt of accounting for the presence of  short-lived nuclei in early solar materials. The first interpretation was advanced by \citet{ct77}, who suggested that a supernova (SN), which occurred close in time and space to the solar nebula formation, might have introduced the required nucleosynthesis contaminations. After this seminal work, the idea of a close encounter with a dying star was re-explored by many authors and various stellar scenarios were examined, from that of a single SN, to the one of a less-massive object, namely a low or intermediate-mass star in its Asymptotic Giant Branch (or AGB) phase 
\citep{pn97,was98,b+99,w+06,gou06,tak08,hus09}. 

Furthermore, also the evolution of the ejecta from massive stars was modelled by various authors. These works ranged from studies of cosmic ray processes inside supernova remnants, where even $^{10}$Be can be produced \citep{t+14}, to the reconstruction  of the evolution of a molecular cloud possibly hosting the solar nebula, with its complex phenomena of nucleosynthesis from previous supernovae, mixing, and contributions from very massive stars in their Wolf-Rayet (WR) stage \citep[see e.g.][and references therein]{gm12,d+17,dw+18}.

Various objections were raised to each of the models presented \citep[for an inventory of them see e.g.][]{b11,b18}. Basically, the most relevant ones can be divided in two categories, each affecting a different mass range of the proposed source. First of all, any idea involving a SN event, or a sequence of SN contributions in a molecular cloud, must keep in mind that these explosive phenomena are the main source of nucleosynthesis for most of the stable elements, and for $\alpha$-rich ones in particular. As suggested by \citet{was98}, materials coming from supernova nucleosynthesis and diluted sufficiently to account for the \alb and \feb concentrations in the ESS are bound to contain large amounts  of oxygen, neon, magnesium, silicon, etc. Their inclusion would affect the early solar nebula in completely different ways, depending on how they are added to the materials of the forming star. Let us consider in more detail this crucial point. A first possibility is that fresh Core Collapse Supernova (CCSN) ejecta are directly added to the forming Solar System by a single event (maybe also triggering the collapse), maintaining the typical, heterogeneous structure characterizing supernova remnants \citep{boss17,loll}. This non-homogeneous and clumpy kind of mixing was early modelled by \citet{pan}; in this case, the effects of the pollution should be now registered in pristine meteorites in the form of wide-spread anomalies in the abundance ratios of stable isotopes for such elements, at the level of a few percent \citep{was98}. The most important of these effects would concern oxygen, generally produced in large quantities by exploding stars leaving a neutron star as a remnant. However, observed oxygen isotope anomalies in primitive Solar System materials do not require the admixture of a distinct nucleosynthetic component (see Section \ref{sec:meas}). Anomalies in contrast with observations would also be predicted for the isotopes of other $\alpha$-rich nuclei and an important excess would be foreseen for the SLR \mn, whose ratio to $^{55}$Mn  would be larger than observed by orders of magnitude \citep{was98,mc00}; the whole scenario is therefore in doubt. Attempts at answering to these objections invoked very peculiar SN types, e.g. objects ejecting only external layers, not interested by manganese production \citep{mc00}, and/or having a very particular mass (close to the lowest limit for CCSNe), to minimize the production of unwanted anomalies on oxygen and $\alpha$-rich nuclei \citep{b+16}. In any case, whatever  ad-hoc choice of parameters is adopted, these models do not really avoid the problems outlined above. An example is the recent work by \citet{b+16}, where abundance shifts not acceptable by the present meteoritic measurements (i.e. at levels of 1-3 \%) remain on various stable isotopes (see their Table 3 in the Supplement materials).

A very different scenario considers that, before being admixed into the presolar nebula, the ejecta of previous supernovae had time to mix and homogenize in the molecular cloud from which the Sun formed, presumably because they came from an appreciable number of explosive events in a sequential star formation process lasting for several million years. If this was the case for the solar formation, the addition of most stable nuclei would not cause any particular problem: they would simply modify slightly the \oq average\cqb solar abundances, not introducing measurable isotopic variations with respect to the average composition in pristine meteorites. This type of scenario has been rather common in recent years \citep{gm12,d+17}. On one side, this scheme might appear rather realistic and is indeed interesting in itself, offering an alternative to the single-star contamination. Some versions of it indicate that a late WR star may have provided adequate amounts of  $^{26}$Al (and maybe $^{60}$Fe) from stellar winds. The presence of other SLRs with different lifetimes would be simply the fossilized record of sequential star formation within a hierarchical interstellar medium. However, as we shall see, a major problem   
in this case emerges for the excess of unstable $^{53}$Mn. While for a single event one might accept that the contaminant was a rare, peculiar SN, where Mn was not ejected, at the level of a whole molecular cloud this idea appears as untenable. We must notice that this kind of problems was not addressed in the papers invoking this explanation, which also did not include the effects of explosive nucleosynthesis \citep[see e.g.][]{gm12} and did not consider  other crucial SLRs, like \csb and \ca. The very short-lived \cab in particular would be completely extinct, in contrast with the striking correlation with \alb shown in Figure 1, which says it must have a stellar origin. Furthermore, these models assumed that the WR star itself does not give significant further contributions to nucleosynthesis, in addition to those from the wind. This is not supported by other works \citep{higdon}. In any case, all the results on addition of SN debris, including ours, are somehow biased by oversimplifications in the treatments of the mechanism for the injection of fresh material, whose consequences still deserve detailed hydrodynamical studies of the type attempted by \citet{boss17} and by \citet{d+17}, but extended to all the measured SLRs.

On the other hand, the alternative origin in a (longer living) AGB star, early suggested by \citet{w+94,w+95} for several radioactivities has the weakness of requiring a chance stellar encounter at a level of probability that borders zero \citep{km94}.

Despite the criticisms, it is clear that a considerable number of radioactive nuclei among those of Table 1 must be of stellar origin and cannot be attributed to endogenic phenomena in the solar nebula itself, like e.g. spallation processes induced by the solar wind \citep{sossi}. In this respect, it was remarkable that the original suggestions by \citet{w+94} on a possible AGB star origin, later specified by \citet{b+03,w+06}, were subsequently and  independently confirmed by \citet{tr09} on the basis of models for slightly more massive stars, hosting hot bottom burning (HBB), i.e. hydrogen burning directly at the base of the convective envelope. In both these approaches it was assumed (with no real proof) that the parent star could produce elements from slow neutron captures (the $s$-process) only through the \neanb source, avoiding the formation of the complementary and rather efficient source \ctan. Even the AGB models, therefore, did not come without drawbacks that go beyond their implausibility. They were however not a priori excluded only because of their apparent success in accounting contemporarily for several SLRs (\al, \fe, \pd, \cs,  \hf, \pb).

In recent years, the abundance measurements of stable isotopes in pristine meteorites were enormously improved, so that any possible scenario for a supernova origin of SLRs must now face more stringent constraints than envisaged by \citet{was98}. On the other hand, also stellar models for both massive stars and AGB giants underwent important improvements. We have now safe predictions of nucleosynthesis from CCSNe, which include rotation in the hydrostatic phases and a full computation of explosive nucleosynthesis \citep{cl13,cl15,cl18}. 
For smaller masses ($M \lesssim 8 - 9$ \ms) attempts have been presented to address quantitatively, on the basis of known physical principles, the mixing of protons from the envelope into the He-shell that is preliminary to the activation of the reaction \ctanb and then to $s$-processing. These new  kinds of models will be used here, in the attempt of limiting the free parameterizations and obtaining more secure indications. The same will be done for massive stars, adopting the mentioned computations where rotation and explosive phases were included. On these bases we shall also comment on the  scenarinos for the evolution of the presolar molecular cloud contaminated by a series of SN explosions and/or WR stars. 

Before performing such a reanalysis, in Section \ref{sec:meas} we present an update of the constraints that astrophysical models must  face, in terms of limits to the isotopic anomalies that nucleosynthesis events can introduce into the solar nebula, without violating accepted measurements. Subsequently, Section \ref{sec:gal} briefly outlines the contributions to SLRs qualitatively expected to come from the uniform evolution of the Galactic disk before the formation of the Sun, thus showing that such a mechanism might
account for most nuclei with a lifetime larger than about 5 Myr. Here we shall make some distinction between nuclei coming from relatively understood phases of stellar evolution and others produced in the very complex and still incompletely
known process of fast neutron captures. We shall then outline a minimal summary 
for (some of) the complexities of this last process in Section 4, trying to make 
clear what kind of characteristics are needed to account for ESS measurements. In particular, in Section 4.2 we show how certain NSM models, confirmed by some recent observations, might offer a way out for the longstanding problems associated with
the ESS abundances of SLRs, especially in the case of nuclei like $^{129}$I and $^{182}$Hf. Subsequently, Section 5 illustrates the possible contributions from a late AGB star polluting the ESS. This is done with reference to recent models for partial mixing connected with the production of both \alb and the neutron source \ctb for $s$-processing. We focus in particular on two recent suggestions for the physical causes of these mixing processes and, for each of them, we show the  predictions for the ESS radioactivities (in Subsections 5.1 and 5.2). There, we underline problems that were not yet clear at the moment of previous publications in this field. Subsequently, Section 6 examines the situation for a late contamination by a Massive Star (MS), also briefly commenting on models of the sequential pollution of a presolar molecular cloud, which call in different ways for a role by MSs. Even in this case, the risks of meeting unsolvable problems is underlined. Finally, Section 7 draws some preliminary conclusions, discussing the parameter space that remains to be explored for attempting an explanation of the very puzzling problem of extinct Solar System radioactivities. 

\section{Constraints from Isotopic Anomalies in Meteorites}
\label{sec:meas}

As was indicated several years ago \citep{was98}, a supernova origin for SLRs present alive in the ESS would necessarily introduce in the parent nebula also variations over the pre-existing record of abundances for stable isotopes. In case the mixing is not homogeneous, this might imply prediction of unobserved isotopic anomalies on elements typically produced by CCSNe. Any model of the solar contamination in SLRs by massive stars ending their evolution as CCSNe with inhomogeneous and clumpy ejecta must come with the
guarantee that the isotopic abundance shifts introduced on the stable isotopes of major elements remain at a level low enough 
not to be in conflict with actual measurements. 

In ancient meteorites, variations in the isotopic composition of oxygen (the most important 
product of SN nucleosynthesis)  are rather large. However, they are, most likely, mainly the result 
of chemical processes and self-shielding in the solar nebula \citep{clay03}, and not the 
fingerprints of distinct nucleosynthetic components. More specifically, O-isotopic ratios 
of bulk chondrites vary by about 10\mbox{\textperthousand}/amu. Could this have an astrophysical origin, e.g. be the result of different abundances of presolar grains? 
The answer is no, as the most primitive meteorites have abundances of presolar O-rich grains of up 
to 500 ppm which, with the typical $^{17}$O-enrichments of a factor of 2 in presolar grains, would 
shift the $^{17}$O/$^{16}$O ratio by only 0.5 permil on a bulk scale.
 
Larger anomalies as compared to those in bulk chondrites are seen in specific components; notably, the 
most extreme cases are CAIs, with their enrichments in $^{16}$O of up to 5\%, as similarly inferred 
for the Sun by the Genesis mission \citep{mc+11} and for the so-called cosmic symplectite (formerly 
known as \oq new-PCP\cq), which shows enrichments in $^{17}$O and $^{18}$O of up to 20\% and 
which is assumed to represent primordial water in the solar nebula \citep{sak+07}. The Genesis data 
suggest that CAIs have inherited mainly the O-isotopic composition of the gas in the solar nebula. 
Mixing the $^{16}$O-rich gas of the ESS with $^{16}$O-poor primordial water components in various 
proportions, along with mass fractionation effects, could easily  account for the variations in 
O-isotopic compositions of planetary materials and there is no need to invoke a distinct 
nucleosynthetic component.

Much smaller isotopic variations are seen for the heavier elements  on a bulk meteorite (planetary) scale, some of which may be of nucleosynthetic origin. For the rock-forming elements Mg, Si, and Fe isotopic anomalies are only at the sub-permil level \citep{teng,poit, dau+}. The same holds for many other of the heavy elements on a bulk meteoritic scale \citep{DS16}. 
Interestingly, relatively large Si-isotopic anomalies (with large experimental uncertainties) were found by the Rosetta mission for the refractory Si component in comet 67P/Churyumov-Gerasimenko, which has 
$\delta$($^{29}$Si) $ = (-145~\pm$ 98)\mbox{\textperthousand} and $\delta$($^{30}$Si) $ = (-214~\pm$  115)\mbox{\textperthousand}. Note however that here errors are at 1-$\sigma$, so that within 2-$\sigma$ the composition would be normal \citep{rubin}. 

In conclusion, on a bulk meteorite scale there is no unambiguous evidence for isotopic anomalies of nucleosynthetic 
origin in excess of a permil. Of course, if we include CAIs, and especially FUN (and hibonite) inclusions, 
as references, things get much more complicated. A useful compilation of isotope data for conventional CAIs and FUN 
(and hibonite) inclusions can be found in \citet{DS16}. 
For CAIs, isotopic anomalies of likely nucleosynthetic origin may reach up to a few permil {\it for certain isotopes}, 
and for FUN (and hibonite) inclusions anomalies can be even much larger, in excess of a percent. However, we are looking for widespread, global signatures, while FUN inclusions are rare and exhibit large mass fractionation effects.
We believe they can be ignored in the present context, and we can consider only conventional CAIs along with meteoritic bulk 
compositions. For them, isotopic anomalies of putative nucleosynthetic origin are clearly much lower on this scale.

As a conclusion, we must verify that, in case the (typically clumpy and inhomogeneous) CCSN ejecta are assumed to be the source of SLRs in the solar nebula, the predicted shifts on the abundances of stable isotopes remain safely below a level of a few permil. Only when the above constraints are verified, one can consider nucleosynthesis processes in a close-by star as being a possible origin for radioactive nuclei in the protosolar cloud.

\section{Contributions from Galactic Evolution}
\label{sec:gal}

For a zero-order estimate of the contributions to SLRs from Galactic evolution, we consider the schematic model of an ISM behaving as a closed-box, enriched over a time duration T, following the approach by \citet{w+06} and by \citet{lug14}. Then, the inventory of a radioactive isotope $R$ relative to a stable nuclide $S$ produced in the same astrophysical site, at the moment in which production previous to the Solar System formation ceases, is:
\begin{equation}
\left[{N^R(T)/N^S(T)}\right]^0_{CE} \simeq \frac{P^R p(T)\tau_R}{P^S<p>T}
\label{UP}
\end{equation}
where CE means \oq Chemical Evolution\cq, the suffix \oq 0\cqb indicates that the estimate is for the moment in which nucleosynthesis ceases, $P^S<p>~$ is the $average$ stellar production rate of the isotope $S$ over the time interval $T$ and $P^R p(T)$ is the production rate of $R$ at the moment when the process ends. Whenever $p(T)$ can be considered as constant ($\simeq <p>$), calling $\Delta$ the delay from the last nucleosynthesis episode after which the Sun forms, one has:
\begin{equation}
\left[{N^R(T)/N^S(T)}\right]^{\Delta}_{CE} \simeq \frac{P^R}{P^S}\cdot{\frac{\tau_R}{T}}\cdot e^{-\Delta/\tau_R}
\label{UP1}
\end{equation}

A problem with this treatment is that we need to apply it to nuclei produced by heterogeneous sources, e.g. by hydrostatic and explosive processes in stars of different mass and by slow and rapid neutron captures. For these last, \citet{wbg} and \citet{w+06} started simply from the assumption of the existence of a unique explosive scenario capable of reproducing the Solar System distribution of $r$-nuclei and derived the production ratios accordingly.  Although in equation 2 we now need only such ratios ($P^R/P^S$) for isotopes of the same element that are very close in mass, the problem of connecting the data for nuclides having different origins (like e.g. $^{53}$Mn, the $p$-nucleus $^{146}$Sm, the $s$-process nucleus $^{205}$Pb or the $r$-nucleus $^{247}$Cm) remains, so that the results of the Galactic enrichment must be considered with a lot of caution and are affected by intrinsic strong uncertainties.

Should one accept the indications by \citet{w+06}, one would find that the ratio $P^R/P^S$ is close to 1 for $^{107}$Pd, $^{129}$I and $^{135}$Cs (the choices were 0.66, 1.3 and 0.724, respectively). This ratio is much smaller ($\simeq 1.4 \times 10^{-3}$) for $^{41}$Ca. Also for \alb the production ratio to $^{27}$Al is low. For example, in supernovae and massive stars $P^R/P^S$ is expected to be between 10$^{-2}$ and 10$^{-3}$: the adopted average value was then 5.4$\times$10$^{-3}$. This last estimate might be suitable to explain the Galactic inventory of \alb  \citep[2.8 $\pm$ 0.8 \ms, see e.g.][]{diehl}. This  corresponds to an average ratio \al/$^{27}$Al of a few $\times 10^{-6}$, which is 10 times smaller than for the ESS \citep[see also][]{higdon}. For $^{60}$Fe, the adopted production ratio to $^{56}$Fe was 2.27$\times$10$^{-3}$ and for Hf, $P^{182}/P^{180} \simeq$ 0.346. Table 2 gives a synthetic view of the abundance ratios that can be obtained with these hypotheses, either at the moment when nucleosynthesis episodes preceding the solar formation ceased ($\Delta = $ 0), or after a delay of the order of the isolation times of cloud cores in star formation regions ($\Delta$ up to $1 - 2 \times 10^7$ yr). 

\begin{table*}[t!!]
\begin{center}
{
{Table 2. SLRs as synthesized by a uniform production model over $T = 10^{10}$ yr} \\ 
{of Galactic evolution. Production factors from \citet{w+06}}

\vspace{0.2cm}
\begin{tabular}{c c c c c c c c}
\hline
\hline
  Rad. & Ref. & $\tau_R$ (Myr) &$[{P^R}/{P^S}]_{CE}$ & $[N^R/N^S]_0$ & $[N^R/N^S]_{10}$ & $[N^R/N^S]_{20}$ & $[N^R/N^S]_{Meas.}$   \\
\hline 
  $^{26}$Al & $^{27}$Al  & 1.03 & 5.4$\cdot$10$^{-3}$ & 5.6$\cdot 10^{-7}$ & -- & -- & (5.23 $\pm$ {0.13})$\cdot$10$^{-5}$  \\
  $^{41}$Ca & $^{40}$Ca  & 0.15 &1.4$\cdot$10$^{-3}$ & 2.2$\cdot 10^{-8}$ & -- & -- &4$\cdot$10$^{-9}$   \\
    $^{53}$Mn & $^{55}$Mn  & 5.3 & 0.189 & 1.0$\cdot10^{-4}$ & 1.5$\cdot 10^{-5}$ & 2.2$\cdot 10^{-6}$ & 10$^{-6}$ \\
  $^{60}$Fe & $^{56}$Fe  & 3.75 &2.3$\cdot$10$^{-3}$ & 8.0$\cdot 10^{-7}$ &5.6$\cdot$10$^{-8}$ & 3.8$\cdot$10$^{-9}$ & 10$^{-8} - 10^{-6}$ \\
  $^{107}$Pd & $^{108}$Pd & 9.4 & 0.66 &6.2$\cdot 10^{-4}$ & 2.1$\cdot 10^{-4}$ & 7.4$\cdot$10$^{-5}$ & (5.9 $\pm$ 2.2)$\cdot$10$^{-5}$   \\
  $^{129}$I & $^{127}$I & 23 & 1.30 & 3.0$\cdot 10^{-3}$ & 1.9$\cdot$10$^{-3}$& 1.3$\cdot$10$^{-3}$ & 10$^{-4}$   \\
  $^{135}$Cs & $^{133}$Cs  & 3.3 &0.724 &2.1$\cdot 10^{-4}$ &1.0$\cdot$10$^{-5}$ &  4.9$\cdot$10$^{-7}$ &  4.8$\cdot$10$^{-4}$\\
  $^{146}$Sm & $^{144}$Sm  & 148 & 0.675 &9.9$\cdot 10^{-3}$ & 9.2$\cdot 10^{-3}$ & 8.6$\cdot 10^{-3}$ &  10$^{-2}$\\
  $^{182}$Hf & $^{180}$Hf  &12.8& 0.346 &4.5$\cdot 10^{-4}$ & 2.1$\cdot$10$^{-4}$ & 9.4$\cdot$10$^{-5}$ & (9.81 $\pm$ 0.41)$\cdot$10$^{-5}$ \\
  $^{247}$Cm &  $^{235}$U & 23 & 3.95 & 8.9$\cdot 10^{-3}$ & 5.8$\cdot$10$^{-3}$ & 3.7$\cdot$10$^{-3}$ & $(1.1 - 2.4)\cdot$10$^{-3}$\\
  $^{205}$Pb &  $^{204}$Pb & 22 & 1.05 &2.3$\cdot 10^{-3}$ & 1.5$\cdot 10^{-3}$  &9.3$\cdot$10$^{-4}$ & 10$^{-3}$\\
\hline
\end{tabular}
}
\end{center}
\end{table*}

The above general picture has not (and never had) the ambition of being really quantitative, both for the uncertainties in the production factors in stars and for the extremely elementary scheme of Chemical Evoution adopted for the Galaxy. It is only a general qualitative view to be improved by future models; in this respect, we cannot aim (in such a rough picture) to obtain an agreement with the measurements at levels better than a factor of 2$-$3; this is a minimum estimate for the uncertainty in the scheme adopted. Especially for Galactic Evolution one should actually consider more sophisticated models, like e.g. in \citet{m+14,boj1,d+17,bm18}

Nevertheless, the above picture already provides relevant pieces of information. It turns out that nuclei of very different origins like $^{53}$Mn, $^{107}$Pd, $^{146}$Sm, $^{182}$Hf, $^{205}$Pb and $^{247}$Cm might actually find a proper explanation for an isolation time between 10 and 20 Myr. Even the very uncertain $^{60}$Fe might not be a problem, in case the most recent estimates for its abundance were to be confirmed \citep{fe60}. The nuclides that are clearly underproduced by this simple and expected process of gradual Galactic enrichment are only limited to \al, \ca, \csb and possibly \feb (this last only in case its initial ESS ratio to $^{56}$Fe should turn out to be larger than 10$^{-7}$). 

However, in the case of heavy n-rich SLRs, the situation is more complex, as pointed out since the beginning by \citet{cam} and \citet{wbg}. Here one has to consider, aside to the $s$-process nuclide $^{205}$Pb, also isotopes of possibly heterogeneous origin, like $^{107}$Pd; and $^{182}$Hf, $^{247}$Cm and others including $^{129}$I,  due to the $r$-process. While the stellar yields of slow neutron-capture nuclei are rather well understood, the situation is quite different for $r$-process isotopes, whose origin is not yet quantitatively established. SLRs make clear that ascribing it to a unique mechanism, taking place in some explosive event of non-specified nature, as done in Table 2, implies to 
find enormous overproductions for $^{129}$I with respect to $^{107}$Pd, $^{182}$Hf and $^{247}$Cm; hence the works by \citet{cam} and \citet{wbg} simply noticed that such a unique scenario was unlikely. Understanding heavy SLR abundances now, some 25 years later, requires to place them in the broader context of more recent observations and models for the $r$-process. We shall try to discuss these issues in the next Section.

\section{Constraints and models for $r$-process nucleosynthesis}

\subsection{Observed constraints and $r$-process sources}

In the $average$ solar system abundances, the decay daughters of $^{129}$I and of $^{182}$Hf have a ratio $N$($^{129}$Xe)/$N$($^{182}$W) = 41.6 \citep{lod}; as $^{182}$W is of $r$-process origin only for about 50\% \citep{tri16}, the abundance ratio between the $r$-components of these nuclei should be slightly larger than 80. Should we adopt the $s$ and $r$ components from \citet{bis}, one would get a higher estimated ratio of 114. On the contrary, in the ESS, given the fact that the isotopic ratios $^{129}$I/$^{127}$I and $^{182}$Hf/$^{180}$Hf are essentially equal ($\simeq 10^{-4}$), the ratio between the two SLRs roughly equals the one of the stable references, i.e. $^{127}$I/$^{180}$Hf, which is about 20. There is a discrepancy by about a factor 4. With respect to the rough predictions of Table 2, referring to the continuous Galactic production of $r$-process nuclei from hypothetical sources of a unique nature, the discrepancy reaches up to a factor of 7 to 10 (see columns 6 and 7). This sharp contrast might probably be accounted for only if the two cases (average  Solar System materials and anomalies in early solids) derive from different origins or different admixtures of $r$-process \oq components\cq. In particular, the average solar-system abundances were built through an elaborated blend of different processes, each accounting for one such   \oq component\cq of the distribution. This blend was established by Galactic evolution over a time scale of $\simeq$ 10 Gyr. SLRs in the ESS produced through fast neutron captures might instead put in evidence the granularity of the Galactic mechanism on shorter time scales, possibly being controlled by only few contributions from specific sources \citep{cam}.

The work by \citet{wbg} tried to infer the origins of the above contributions; although at that moment the reference sources were mainly assumed to be CCSNe, through neuton captures occurring in a neutrino-driven wind \citep{wos94}, this assumption actually does not enter directly in the estimates of Table 2, which simply require a single mechanism reproducing the solar $r$-process abundances. The same approach was discussed by \citet{b+99}. As a simple recipe for finding a way out, \citet{wbg} guessed that the astrophysical source for the production of $^{129}$I appeared for the last time in the solar neighborhood a long time before the last event producing $^{182}$Hf and $^{247}$Cm, thus implying a much longer decay of the first one, from which the low ESS $^{129}$I abundance would derive. This was subsequently extrapolated by \citet{qw00}, who assumed more explicitly a specific source for these nuclei (CCSNe) and proposed that two types of them were at play, one producing the lighter $r$-nuclei, up to $A \simeq $ 130 (including iodine) and the second producing the heavier nuclei. The first kind of events was indicated with the letter $L$ ($lower$, from their low expected rate of occurrence) and the second was indicated with $H$ ($higher$). $^{107}$Pd would be low in that hypothesis, but for it the help of the $s$-process can be invoked, see Section 5.

The above ideas generated extended debates and were in general criticized as being too simplistic. It was in particular underlined that, within CCSN models for the $r$-process, any {\it physically based} mechanism yielding enough $^{182}$Hf  to explain its ESS abundance, would most probably produce similarly also the nuclei at the $N = 82$ peak, including $^{129}$I, thus leading to high values of their ratio \citep[see e.g.][]{pok}. In other words, the requirement of having \oq pure\cqb $r$-process sources separately producing the two nuclei seemed to be too ad-hoc to be accepted.
More plausible might be situations where they are produced together but at different efficiencies. This would be for example the case, using a rather recent mass model like the \oq finite-range droplet model\cqb update of 2012, or FRDM(2012), by \citet{mo12}, if a weakening of the $N = 82$ shell closure (often referred to as \oq shell quenching\cq) were to occur below the doubly-magic $^{132}$Sn \citep{dill,AAB}. This would anticipate the maximum of the peak, perhaps down to $A = 126$ \citep[see Table 1 in][]{kfhpo}. Then $^{129}$I would not stay at the peak, but after it, with a reduction of its abundance by a factor of 2$-$3 with respect to a standard solar $r$-component \citep{far}. 

Subsequent research then clarified that other sources, different from CCSNe, might be crucial in producing the $r$-process nuclei \citep{frei}. At the moment of this report, 
two main sites are often discussed to have excellent chances to contribute, in various proportions, to produce nuclei from fast neutron captures. They are the rare magneto-rotational types of supernova \citep[MRS; see e.g.][and references therein]{nishi17} and neutron star merger (NSM) events; \citep[see e.g.][]{frei}. One has in particular to mention that this last paradigm received a now publicly famous observational support in the recent gravitational-wave event GW170817 \citep{ligo}. In the electromagnetic source AT2017gfo, the kilonova associated to it, qualitative evidence was recorded of the production of heavy $n$-rich elements at the second and/or third abundance peak \citep{pian17,tan17,t+17}. The possibility of having $r$-process nucleosynthesis in such an environment was early suggested by \citet{lat77}, \citet{mey89}, and \citet{eich89}. This kind of models was recently indicated to be in principle able to explain the whole solar-system distribution of nuclei coming from fast neutron captures \citep{wan14,thiel1}.
One has also to consider cautiously that the traditional CCSN models seem not to be
completely out of the picture yet \citep{km17}. There have also been persisting suspects that CCSNe of a specific type \citep[faint, $^{56}$Ni-poor, type IIP events, showing enhanced Sr II and Ba II lines, like SN2009E and perhaps SN1987A, see e.g.][]{pasto}, can contribute to the $r$-process, as early suggested by \citet{ts01}. Although the strength of Ba II lines might be affected by variable ionization and temperature issues \citep{uc5,pasto}, these effects should be valid in general and might not explain completely why in other CCSNe the 
Ba and Sr lines are significantly weaker \citep{bw17}. The products of the above faint SN II sources might have been observed in dwarf galaxies \citep{ji16}. In general,
the real relative importance of the various contributors to heavy neutron capture nuclei, their frequency of occurrence and the amount of processed material returned to the Galaxy remain still unsecure, although NSMs, where high neutron excesses are found, are emerging as one of the most promising sites for $r$-processing \citep{thiel18}. 

Constraints on the zoo of present-day models can be found by considering real observed or measured element admixtures \citep{kr93}. In our case, these include not only the solar system average abundances and ESS radioactivities, but also the pattern of $r$-nuclei
observed at low metallicity, in our Galaxy and its neighbours, when the $s$-process has not yet started to appear and one has chances to see stars polluted by only part of the long-term $r$-process blend \citep{SCG}.
In this framework, it is by now ascertained that many old stars exist showing an almost solar distribution of elements across and beyond the $N = 82$ magic neutron number \citep{spite}. The prototype of these objects is the famous CS22892-0052 source \citep{s+96}, a supergiant in Aquarius. When in these stars one considers also lighter species (with $Z$ = 40 to 50) one sees that for them the scatter is large, but on average the production of these elements is lower by a factor of 2-3 with respect to a scaled solar $r$-process distribution \citep[see e.g.][in particular their Figure 5]{hon}. Completely different metal poor stars however exist, at even lower metallicity, the prototype being the subgiant/giant star HD 122563 \citep{hon}. Here, the light $r$-nuclei at $N = 50$ are dominant with respect to heavier species across and beyond the $N = 82$ peak. We further note that the inventory of low-metallicity stars differently enriched in the $light$ and $heavy$  $r$-nuclei seems to suggest that the former ones are actually more frequent than the latter ones, but also that the amount of $r$-processed matter ejected by events producing the lighter $r$-process elements be much smaller than for those producing preferentially the heavier nuclei, maybe by 2 orders of magnitude \citep{macias}. The two producing environments might be of heterogeneous origin (e.g. CCSNe and NSMs) or of the same type (e.g. NSMs only), but occurring in different conditions. For example it has been shown that NSM phenomena can give rise to light or heavy $r$-nuclei, depending on the extremely variable possible conditions. Anyhow, deriving absolute production factors like the $<p>$ values needed in Equation 1 is certainly premature, so that we are obliged to stay at a purely qualitative discussion. Moreover, both HD 122563 and CS22892-0052 reflect situations probably not suited to explain the ESS $^{129}$I/$^{182}$Hf ratio. Although the most relevant elements I, Xe ($Z$ = 53-54) and Hf, W ($Z$ = 72-74) were not observed, looking in general at the closest elements we see that, from both cases,  we would expect a ratio much higher than observed in the ESS.  Hence, if this last sample reflects the isotopic ratios typical of a specific $r$-process variety, then the two types of metal-poor stars (albeit offering a closer and different look at the granularity of the process) should be already the products of different forms of admixtures of heterogeneous components. 

Although with caution, we can say that something better, more similar to the
individual $r$-component revealed by ESS heavy SLRs, actually may exist, in some metal-poor stars. A few years ago it was shown, in \citet{ro16}, that three out of four stars observed in the dwarf Galaxy Reticulum 2, although in general similar to the Sneden's star, actually show rather large abundance ratios between the heavy elements before the third $r$-process peak (e.g. Dy) and those immediately after the second peak (e.g. Ba). In the mentioned stars the ratio Dy/Ba ranges from 7 to less than 24 (upper limit) times the average solar value. This last needs to be corrected for deriving a pure (solar) $r$-process component. This can be done using the $r$-residuals (1-$s$) from the models quoted in the next Section, in particular from \citet{pal17b}. With respect to previous computations, these models yield lower estimates for the Dy$_r$/Ba$_r$ ratio in the Sun, in the range 2.7 to 5, with an average value depending on the Initial Mass Function adopted for the weighting, on mass loss rates etc. One can roughly evaluate it to be near 3.5 \citep[against a previous estimate of 5.7, derived from][]{bis}. We underline that the new estimates are quite uncertain; despite this, they are closer to what can be found in HEW models with shell quenching \citep[see e.g.][]{far}, where values down to 2.0 $-$ 2.3 can be be obtained. With the above correction (on average by a factor of 3.5) for the solar $r$-component normalization, the Dy/Ba ratios in the three stars of Reticulum 2 become 2, 4.5 and less than 6.8 (upper limit) times higher than in the solar $r$-process distribution.  Can one roughly assume that this implies relatively large ratios also near the (unobserved) W and Xe (whose isotopes at $A$ = 182 and $A$ = 129 are decay daughters of the SLRs we are discussing)? As we illustrated, we need an enhancement factor in $^{182}$Hf/$^{129}$I of about 4; if our extrapolation is correct, the observations of Reticulum 2 early showed real stars where this might be achieved. A similar situation may apply to the more numerous stars recently observed by \citet{hans1}. In their Table 5, several measurements identify objects (called $r$-I) with negative values of [Eu/Ba]; some of them yield linear Ba/Eu ratios lower than in the solar $r$-distribution. Models accounting for this trend include the mentioned HEW cases with shell quenching, or certain NSM scenarios, like e.g. those by \citet{gor13}. Hence, two different sites, affected by $r$-process varieties similar to the one producing the heavy SLRs, would have been observationally confirmed. At present, NSMs seem to be the most probale sources to explain the above abundance distributions, due to their lower $Y_e$ values with respect to CCSNe. 

Anyhow, the abundance distributions of different metal-poor stars confirm that various $r$-process varieties must necessarily exist and that the blend shown by the average solar composition is certainly not \oq universal\cq, as also indicated by the extensive theoretical work of the last 25 years \citep[see e.g.][and references therein]{roede,thiel18,tsu18,km17}.

With the above scenario in mind, one can look for parameter studies, based both on site-independent and on site-specific models, to identify the astrophysical conditions requested for reproducing the observational evidence. From these conditions we can then try to figure out plausible scenarios accounting for the abundances of $^{129}$I and $^{182}$Hf in ESS samples, without violating other constraints from low-metallicity stars.

\subsection{Reconciling the ESS abundances of $^{129}$I and $^{182}$Hf.}
In general it is recognized that a very promising general scheme for $r$-processing involves neutrino-driven interactions in neutrinospheres and/or neutrino winds (NWs), established in explosive conditions above a neutron star  \citep{frei}. It might be found in various astrophysical scenarios \citep[CCSNe, NSMs, MRSe, see e.g.][]{km17}. 

A few works some years ago criticized earlier attempts aimed at modelling its occurrence in CCSNe \citep[see e.g.][]{frohl,wan12}, inferring that, for those conditions, the NW would remain proton-rich during its entire life, precluding any $r$-process nucleosynthesis, even simply for producing light nuclei up to the first magic neutron number $N = 50$ (Sr, Y, and Zr). However, \citet{rrs12} re-established this scenario as a possible one for the occurrence of neutron captures. The above authors showed that, with a more detailed treatment, including the nucleon potential energies and the collisional broadening of the response, the previous negative conclusions would have been considerably changed. In particular, for a reasonable period of time, the NW was predicted to remain moderately neutron rich. Interactions occurring in the NW were described, e.g., in \citet{ok,far,mp17,thiel18}. 
The main parameters controlling the nucleosynthesis products are the number of neutrons per nucleus $Y_n/Y_{r-seed}$, the number of electrons $Y_e$, the expansion velocity, $v_{exp}$, and the entropy per nucleon, $S$ (generally expressed in units of the Boltzmann's constant, $k_B$). They are linked by the relation \citep[][]{far08a,far08b,kra08}:

\begin{equation}
Y_n/Y_{r-seed} \simeq v_{exp} \cdot \left({S \over Y_e}\right)
\end{equation}

In CCSNe only limited neutron enrichments seem to be achieved ($Y_e = 0.4-0.45$); in order to produce heavy nuclei, this requires that high values of the entropy $S$ per nucleon are available ($S \geq 200$), which fact gives to the mechanism its same denomination: {\it High Entropy Wind}, or HEW. Nucleosynthesis induced by neutrino interactions can occur also in contexts different from CCSNe, in particular in NSMs, where the neutron excess is always much higher (with $Y_e$ values down to 0.2); here, lower values of the entropy $S$ are required ($S \lesssim$ 20, {\it Low Entropy Winds}). These were some of the reasons why so much attention was dedicated to this scenario in recent years.

In the original work by \citet{rrs12} the values of $Y_e$ and the maximum entropy   $S_{Max}$ for CCSNe were such ($Y_e \simeq 0.45, S_{Max} \leq 100$) that only relatively light trans-Fe elements, from Sr to maybe Ru, could be produced, in a process with a high proton abundance. A result of this kind had been previously discussed by 
\citet{far1}, in addressing the composition of presolar SiC grains of type
\oq X\cqb \citep{pellin}.

However, after the indications by \citet{rrs12}, a few groups showed that in some 
scenarios HEW models could still apply, reaching high values of $S_{Max}$ \citep{kfm,tud}. As said, this is a necessary condition for the process to be effective at relatively high $Y_e$ values, as in CCSN contexts. For example, explaining with these sources the observations of extremely metal poor stars like HD 122563 \citep{hon} with ejecta from a moderately neutron-rich wind \citep[$Y_e = 0.45$, as in][]{rrs12}, would require $S_{Max} \simeq 220$. Should one try to account for other metal-poor stars richer in heavy $r$-nuclei, like CS22892-0052 \citep{SCG}, in the same $Y_e = 0.45$ condition, then $S_{Max}$ as high as 280$-$300 would be needed \citep{far,kfm}. If, instead, the game is played in more n-rich environments like NSMs, these requirements would be reduced by roughly a factor of ten \citep{thiel1}. 

One has also to notice that virtually any observed star, even at very low metallicity, contains an admixture of light and heavy $r$-nuclei. Just to make examples, r-poor stars like HD 122563 have Sr/Eu values from about 100 to about 550, whereas r-enriched stars like CS22892-0052 have Sr/Eu ratios from about 20 to about 30. Recent extreme cases were shown by \citet{ji16b} for stars enriched in heavy (A $>$ 130) nuclei; they have Sr/Eu ratios lower than 10 and down to a minimum of about 3.5. If these numbers define a pure \oq main\cqb $r$-component, this means that all other observed stars contain admixtures of different processes, i.e. are characterized by wider blends of $Y_e$ and $S$ values than obtained in individual  calculations \citep{fb18}. This is so even at very low metallicities ([Fe/H] $\lesssim -2.5$), where the stars should have been polluted by only very few SN events. If NSMs are at play, their potentially much lower $Y_e$ values would allow all $r$-nuclei up to the heaviest ones to be produced in rather low entropy conditions \citep{GJ,thiel18}. 

In general, what we expect as a result of neutrino-wind driven nucleosynthesis phenomena for increasing values of $S_{Max}$ can be outlined as follows.

\begin{itemize}
\item In CCSNe, for the lowest values of $S_{Max}$, in the dynamics of the mechanism a primary-like, rapid nucleosynthesis process can start, mainly  controlled by charged particle (CP) interactions, where $Y_n/Y_r$ is lower than unity. In these conditions SiC grains of type \oq X\cqb might find their production site \citep{pellin,far1}. This requires $S_{Max}$ to be up to 100 for typical $Y_e$ values of 0.45 \citep{far1}. This condition is not met in NSM environments, where the material is always neutron rich.

\item 
For increasing values of  $S_{Max}$ and of $Y_n/Y_{r}$, neutrons start to dominate and we find various varieties of rapid n-capture processes. These variants have been called $weak$, $incomplete$, or, for higher $S$ values, $main$ and $actinide-boost$ $r$-processes. Typically, the same regions of $A$ can be reached in NSM models for $S$ values smaller by an order of magnitude, due to the generally much lower $Y_e$.

\item 
In CCSNe, for values of $S_{Max}$ around 150 and in the range of free-neutron abundances 1 $\lesssim Y_n/Y_{r} \lesssim 15$ one would find what was called by stellar modelers the \oq weak\cqb $r$-process. Although light $r$-nuclei are produced, these are not the conditions for explaining the star described by \citet{hon}, as this last includes, as  said, some heavy nuclei, while here the production stops at the rising wing of the $A = 130$ peak, producing iodine at the low level of a few percent. Similar results can be obtained with NSM models at values of $S_{Max}$ at least ten times smaller \citep{sie17}.

\item
There might be a following, more effective, $r$-process mechanism, for larger values of $Y_n/Y_{r}$ and for $S_{Max}$ values ranging between 200 and about 220 in CCSNe (and again smaller by an order of magnitude in NSMs, with $Y_e$ values smaller by typically a factor of 2). Its existence is certified by stars where nuclei in the region from Sr to Cd dominate, but are accompanied by variable abundances of heavier species, like Ba and Eu. Examples of such stars with a rather scattered composition are abundant. Aside to 
the one quoted by \citet{hon}, new measurements have been presented by \citet{hans}: this was called \oq limited $r$-process\cq, repeating suggestions advanced by \citet{roede} and others.

\item A main $n$-capture mechanism follows, responsible for the top of the $A \simeq$ 130 peak, including most of Xe, the Rare Earth Elements and the $A \simeq$ 195 third $r$-peak. This process is often referred to as the \oq Sneden-like\cqb $r$-process, as it would account for stars with a solar-system distribution of elements above $A \geqslant 130$, like those described in \citet{SCG}, plus a variable proportion of lighter nuclei. This kind of process might occur in CCSNe for values of $Y_n/Y_r$ up to about 150 and values of $S_{Max}$ above 220. Again, in NSM scenarios suitable conditions require much smaller values of $S_{Max}$, due to the low $Y_e$. In this process the enhancement factor 
for iodine would traditionally reach up to more than 90\% of the most effectively produced elements, like Eu. However, its abundance can be considerably reduced if 
a weakening of the $N = 82$ shell closure (often referred to as \oq shell quenching\cq) occurs below the doubly-magic $^{132}$Sn \citep{dill,AAB}. This would anticipate the maximum of the peak, perhaps down to $A = 126$
\citep[see Table 1 in][]{kfhpo}. In those conditions the ratio I/Hf would be considerably reduced. The same effect would be obtained also by further increasing $S_{Max}$. Models of the $main$ component with a low I/Hf ratio seem to be possible also in NSM models with high efficiency, like e.g. in \citet{gor13,GJ,bau}.
 
\item There might be a further, limited contribution from a vey efficient $n$-capture process (that would explain the \oq actinide-boost\cqb stars). Some of the stars  observed by \citet{roe09} might have these characteristics.

\end{itemize}

Depending on the ambient conditions, any site might be characterized by a specific range of $S$ values; their superposition, from many different events, gives rise to the robust distribution of $r$-process nuclei observed in young Galactic stars; in the specific case of the $main$ component, nuclei with $A \geqslant $ 135, $Z \geqslant $ 56 are produced with abundance ratios that look remarkably constant, since their early appearance in low metallicity stars \citep[see e.g.][and references therein]{SCG, hill}.


We have then two possible ways out for the I/Hf ratio. On one side, the same $main$ component might produce a much lower abundance of iodine than previously found in CCSN models. This is obtained in certain NSM scenarios producing rather heavy $r$-nuclei \citep{gor13,bau} and might also be found in CCSNe, where very high entropies and shell-quenching effects substantially reduce the previously expected I/Hf predictions. In both cases, the observed ESS ratios might be the direct outcome of only one specific $r$-process site, possibly the same producing the abundance patterns observed in some halo stars by \citet{hans1} and in Reticulum 2 by \citet{ro16}. In view of the fact that NSM models show a high number of free neutrons, this scenario is probably the most promising one for yielding the required $^{129}$I/$^{182}$Hf ratio. Another variety of NSM phenomena, with a lower number of neutrons per seed, might then be at the origin of $^{107}$Pd, in a weaker $r$-process. As mentioned, this $weak$ component might also come from CCSNe in suitable conditions. Table 2 would still be broadly similar to reality, but the production factor $<p>$ for $^{129}$I should be reduced by a large factor, maybe of the order of 5 ($\pm$ a factor of two, trusting the few data from Reticulum 2).

Alternatively, the ESS $^{129}$I might derive from the averaging over time scales much shorter than 10 Gyr of the contributions from different sources; a very limited number of them with a $traditional$  main $r$-mechanism and a much larger number with a $weak$ mechanism, with a poor efficiency (a few percent) in $^{129}$I production.  $^{182}$Hf, instead, would be fully produced by the $main$ $r$-process, with negligible contributions from the sources responsible for weaker components. In this second case, the productions of I and Hf would be essentially decoupled; however, in that case we would have the problem of obtaining the correct relative efficiencies (in frequency and mass ejected) of different producing sites for explaining Pd, I and Hf in an admixture of different sources. This seems a prohibitive fine-tuning task now; we therefore tentatively indicate the first possibility as the more probable.

We mention here that $^{182}$Hf is well accounted for by the Galactic enrichment of $r$-process elements. We therefore do not feel any need to increase its $s$-process fraction, as done in \citet{lug14}, following the revision of the level scheme and of the decay rate of $^{181}$Hf, ensuing by a single  estimate \citep{bond}. On this point we prefer to maintain a cautious approach, waiting for possible confirmations of this individual measurement.

\section{The Reference Models for Intermediate Mass Stars}
\label{sec:AGB}

When a radioactive nucleus is not accounted for by the chemical evolution of the Galaxy and requires a late event of nucleosynthesis to explain its abundance in early solids of the Solar System a formalism slightly different from equation (1) applies to it. As discussed in \citet{w+06}, if a nearby star produced a radioactive nucleus $R$, of mean life $\tau_R$, introducing for it in the ESS an abundance ratio $\alpha^{R,S}$ with respect to a stable isotope $S$ of the same element, the following relation holds between the $\alpha^{R,S}$ value and the abundance ratio $N^R/N^S$ (radioactive versus stable) in the stellar envelope:
\begin{equation}
\alpha^{R,S} = d \cdot \frac{N^R}{N^S}q^Se^{-\tau_R/\Delta}
\label{eq1}
\end{equation}

Here $q^S$ is the enhancement factor of the stable isotope $S$ in the same envelope. The parameter $d$ represents a dilution factor that measures the fraction of the ejected wind that is incorporated into the forming solar cloud, while $\Delta$ has the same meaning as in equation (2).

In considering the stellar sources suitable for a late contamination of the solar nebula with SLRs, in this Section we start with an estimate of the possible role played by an Intermediate Mass Star (IMS). This issue was recently addressed also by \citet{was17}. They assumed that below an initial mass of about 5\msb a \ctb pocket could be formed during the AGB phases, inducing the reaction \ctanb and producing neutrons efficiently through it. Above this limiting mass they instead considered only neutrons from the \neanb reaction. These more massive models were found to experience HBB at the base of the convective envelope, producing efficiently \al. Contrary to previous indications by \citet{tr09}, those authors could not find an explanation for SLRs in their models. In the lower mass range \alb was insufficiently produced with respect to $s$-elements; for the higher masses the reverse was true. They suggested that a compromise solution might be found mid-way between the two cases, but did not present a detailed model for it. One has to notice that in these computations the \ctb source was introduced ad hoc and without a physical model for it, as was common in many computations of the last 20 years or so. Moreover, also for HBB the model-dependency is large, so that these calculations cannot be considered as conclusive. 

The approach we want to follow here is different. Whenever possible, we would like to base our considerations on physical models that avoid (as far as it is possible today) the free parameterizations. This attempt must address first of all the chemical peculiarities of light elements and  CNO isotopes observed in stellar photospheres \citep[see e.g.][]{gil89} 
and not directly accounted for by traditional stellar models. These peculiarities
trace the existence, in stars below 7-8 \ms, of non-convective transport phenomena. In particular, for low mass red giants, several authors \citep[see e.g.][and references cited therein]{bus10,pal17a} have shown that the known episodes of convective mixing that occur after the star enters its Red Giant Branch (RGB) stage and that carry to the envelope materials previously processed by nuclear reactions, are not sufficient to explain their isotopic abundance observations from $^7$Li up to $^{26}$Mg. The most important mixing episodes of this kind are called the \oq first dredge-up\cqb (hereafter FDU), and the \oq third dredge-up\cqb (hereafter TDU). 
The first one is induced by the inward expansion of the envelope after the main sequence; in solar-metallicity stars its main effect is a reduction 
of the \cd/\ctb ratio at the start of the RGB stage to 25-30 (from the initial solar value around 90), and a contemporary increase of the $^{14}$N surface abundance. The second mixing episode is a similar envelope penetration occurring repeatedly, after runaways of the He-burning shell called \oq thermal pulses\cqb (TP), during the final AGB phase. It mixes to the envelope mainly helium, carbon, and $s$-process elements. Stars more massive than about 2.2 \msb also experience a \oq second dredge-up\cqb (SDU) in early phases of the AGB, carrying up materials polluted by extensive H-burning processes, including He and $^{14}$N. In Subsection 5.2 we shall see that further consequences of interest for the present study may also emerge.

For understanding the further mechanisms of mixing and nucleosynthesis that must be at play in evolved low- and intermediate-mass stars, stellar spectroscopic data were crucially supplemented by the record of isotopic abundances accurately measured in presolar grains found in pristine meteorites, as most of these grains were actually formed in the circumstellar envelopes of AGB stars. \citet{wbs} suggested that the peculiar isotopic composition of oxygen found in a large number of presolar corundum (Al$_2$O$_3$) grains could be explained assuming the presence of a deep matter circulation in the mentioned stellar sources; the same process would also be responsible for the presence of $^{26}$Al in some of these grains as well as for the high abundances of \ctb and the spread in the concentration of $^7$Li observed in low mass red giant stars. These suggestions were then confirmed by \citet{nol} and later by \citet{pal11a,pal11b}.

Since then, several works have been presented by various groups to interpret those results, originally obtained with parameterized approaches, on the basis of models physically built on some fundamental properties of stellar plasmas. These last ranged from rotational mixing \citep{cl10} to thermohaline diffusion \citep{egg1,egg2,cz}, to asteroseismic effects inducing gravity waves \citep{dt03} and to the transport guaranteed by the buoyancy of magnetic flux tubes \citep{bus07}. Later, some of these mechanisms were found to be too slow to induce remarkable effects on the RGB (and even more on the shorter AGB phase): this was for example the case with thermohaline diffusion \citep{dm11}. Other mechanisms were not applied in detail to the interpretation  of isotopic abundances on both the RGB and AGB and then compared to constraints coming from presolar grains.

Two remarkable exceptions however exist. On the basis of considerations concerning the physics of the inner border of the convective envelope of a red giant, and subsequently also the effects of rotation, \citet{cri09,cri11,cri15a,cri15b} performed a general revision of the models, both for low mass stars (LMS, M/M$_{\odot} $ $\lesssim$ 2 - 2.2) and for IMS ($2  - 2.2  \lesssim$ M/M$_{\odot} \lesssim 7 - 8$), where the effects of partial mixing were introduced along the whole evolutionary history and were extended to include the inner He-rich layers, where the presence of deep-mixing (DM) phenomena causes the formation of a \ct-reservoir suitable to induce the activation of the \ctanb neutron source. 

Independently of that, \citet{nor08} and subsequently \citet{nb14} showed that the known mechanisms associated with magnetic stellar activity might induce circulations and transport phenomena in the external layer of a star   that in specific circumstances might become quite fast (up to 100 m/s). Along the RGB and AGB sequences, their activation can carry materials modified by nucleosynthesis in H- and He-burning shells to the surface of the star. Detailed calculations of the consequences of such a suggestion were subsequently performed by \citet{tri14,was15,tri16}; more recently, by \citet{pal17a,pal17b}, comparing the results with a series of constraints ranging from the Solar System distribution of $s$-elements, to the record of oxygen isotopic anomalies and of $^{26}$Mg excesses induced by the in-situ decay of $^{26}$Al in presolar oxide grains, and up to the isotopic admixtures of trace $n$-rich elements in presolar SiC grains.

In the following two Subsections we shall briefly review what kind of predictions for SLRs in the early solar nebula can be derived by assuming that the forming Sun was contaminated by the slow winds of an AGB star of intermediate mass, hosting either of the two mentioned deep-mixing processes. The reference models we adopted have similar general input parameters. Those of Subsection 5.1 adopted the compilation by \citep{lod} for scaled solar abundances and that by \cite{dil14} for neutron capture cross sections. For the models of Subsection 5.2 the choices were \citet{lod03} and \cite{bao}; none of the small differences present in these databases has any effect on the results discussed here. In both cases, as mentioned previously, we did not adopt for $^{182}$Hf the suggestions by \citet{lug14}, indicating a revision of the decay rate for the precursor 
$^{181}$Hf, which would substantially increase the $s$-process contribution to $^{182}$Hf.  This suggestion was based on a single measurement and, on the basis of our discussion of Section 3, this last SLR seems already well explained by the Galactic enrichment in $r$-process nuclei. We prefer in these conditions to cautiously wait for new experimental evidence.

\subsection{Effects of a Late AGB Star. I. Models with MHD-Induced Mixing}

\begin{figure}[t!!]                                               
\begin{centering}
\includegraphics[width=0.9\columnwidth]{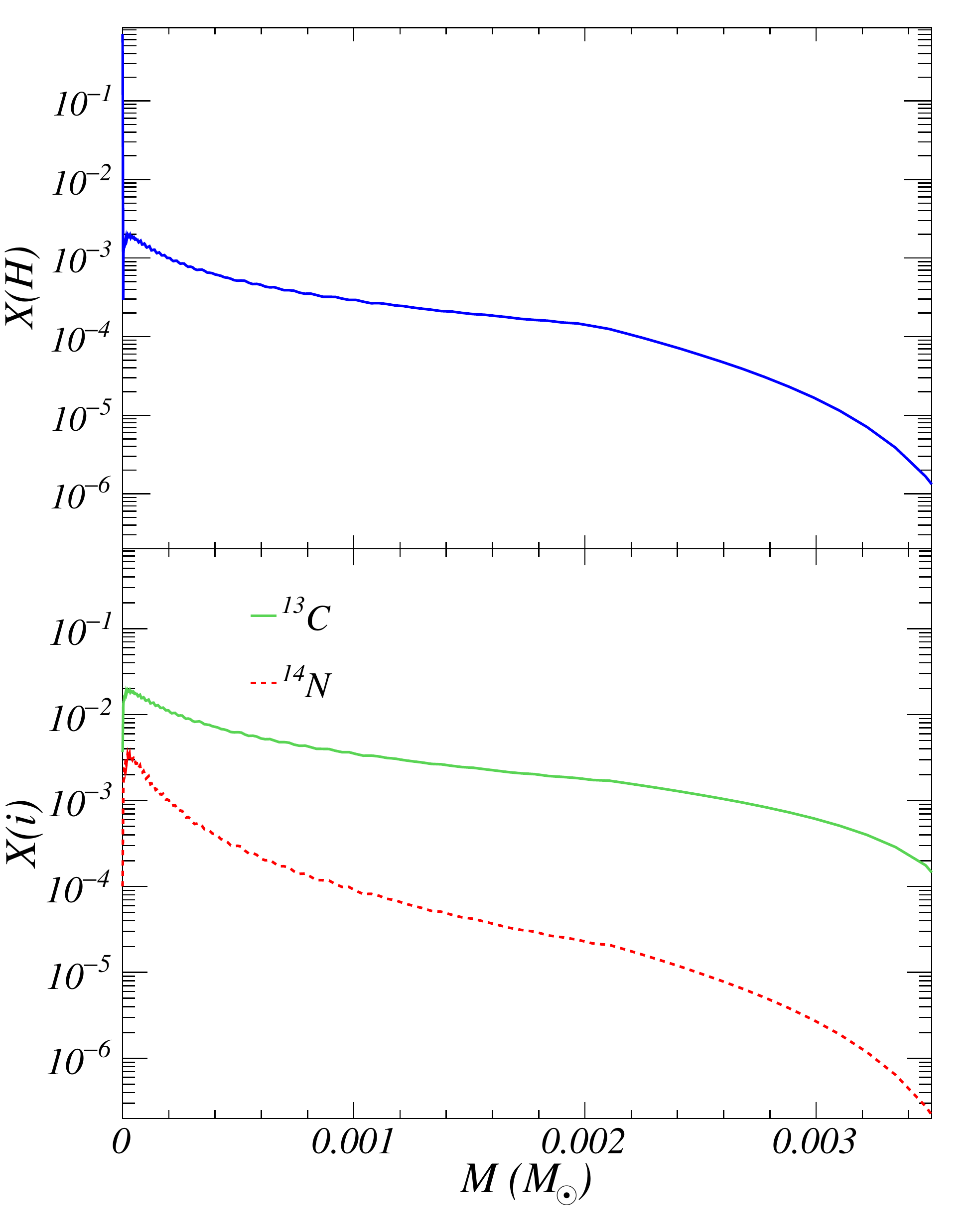}                
\caption{Upper panel: The proton profile established in the He-rich layers as a consequence of mass conservation, when magnetic buoyancy is occurring at a TDU episode. The case represented is for a 3 \msb star of solar metallicity, at the sixth TDU. The extension is about 60\% of what was found by \citet{tri16} for a 1.5 \msb model. Lower panel: The ensuing profiles of \ctb and $^{14}$N formed after the H-shell reignition.}
\end{centering}
\end{figure}

\citet{nb14} demonstrated that the very complex MHD equations valid for a stellar plasma might simplify drastically in some specific geometries well approximating the radiative layers below the convective envelope of an evolved star. In this particular case, the equations can be solved analytically in an exact way, yielding simple formulae that can be introduced into stellar models to mimic the local effects induced by the magnetic field. In particular, the solution refers to the general process of buoyancy of magnetized zones early described by E.N. Parker in the fifties, leading to their emergence in the convective envelopes at a rather fast speed. 

While this simple analytic solution to the MHD equations can be found under rather broad conditions, when we impose that the result must have a physical meaning appliable to stars, one is led to require that a number of constraints are satisfied. They can be summarized as follows:

\begin{enumerate}

\item The density must drop with radius as a power law ($\rho(r) \propto r^{k}$) with an exponent $k$ that is negative and has a modulus larger than unity.

\item Also the pressure must follow a similar trend, but with a slightly larger negative exponent, so that a polytropic relation of the type $P(r) \propto \rho(r)^{\delta}$  with $\delta \lesssim 4/3$ holds.

\item The \textit{magnetic Prandtl number $P_m$} (namely the ratio between the \textit{kinematic} viscosity $\eta =\mu_d/\rho$ (where $\mu_d$ is the $dynamic$ viscosity) and the magnetic diffusivity $\nu_m$) is much larger than unity \citep[see][]{SPI62}. 

\item While the $kinematic$ viscosity $\eta$ cannot be neglected, the $dynamic$ viscosity $\mu_d$ remains small due to the low density (at the level of one to few percent).

\end{enumerate}
 
Once those conditions are verified, the radial velocity of magnetized structures turns out to be:

\begin{equation}
v_r=\Gamma r^{-(k+1)}
\label{eq2}
\end{equation}

where  $\Gamma = v_p r_p^{k+1}$. The parameters $v_p$ and $r_p$ refer to the buoyancy velocity and radial position of the innermost layer where the above-mentioned conditions start to be satisfied, while $k$ is the exponent in the relation $\rho \propto r^k$.                                              

Below a convective envelope $k$ is always rather large in absolute value, but negative ($k \le -3$); then equation (6) yields an unstable condition, in which the buoyancy starts slowly but then gains speed rapidly for increasing radius. The toroidal component of the magnetic field can be correspondingly written as:
\begin{equation}
B_{\varphi} = \Phi(\xi) \left(\frac{r_p}{r}\right)^{k+1}
\label{eq3}
\end{equation}
where $\xi$ is an adimensional variable and $\Phi$($\xi$) can be chosen with a lot of freedom; for a simple solution it might even be a constant ($\Phi$($\xi$) = $B_{\varphi,p}$). Again, we use the suffix \oq $p$\cqb to indicate the values of the parameters pertaining to the layer from which buoyancy (on average) starts. For more details on the solution and its stellar applications see \citet{nb14,tri16,pal17a,pal17b}.

During the occurrence of a TDU episode, while the H-burning shell is extinguished, the above procedure describes a mass upflow that forces a downflow of protons from the envelope for mass conservation.
In the light of the above considerations, the mixing rate forced by magnetic buoyancy is:

\begin{equation}
\dot M = 4 \pi \rho_e r_e^2  v_ef
\label{eq4}
\end{equation}

Here $v_e$ is the velocity of buoyant flux tubes at the envelope bottom and the filling factor $f$ is of the order of 10$^{-5}$ \citep{tri16}. By applying this equation to AGB stars of masses up to 5 \ms, $\dot M$ values in the range 10$^{-7}$ to 10$^{-5}$ M$_{\odot}$/yr can be obtained. 

The downflow of matter from the envelope, pushed down by the rising material, was analyzed in detail by \citet{tri16} for the formation of a $^{13}$C-pocket, and the subsequent  neutron release via the $^{13}$C($\alpha$,n)$^{16}$O reaction. The $^{13}$C reservoir varies in mass (by up to a factor of three) during the sequence of thermal pulses of an individual star. It varies more substantially as a function of the stellar mass.

A typical set of abundance profiles for protons, and subsequently for $^{13}$C and $^{14}$N, as obtained with our model in the He-rich layers is represented in Figure 2. Table 3, instead, shows the extension in mass of the pocket at the sixth TDU episode of different stellar models (the metallicity is indicated in the common logaritmic spectroscopic notation [Fe/H] relative to solar, so that [Fe/H] = 0 means a solar metallicity, [Fe/H] = $-0.5$ means one-third solar). As is shown by the table, the size of the pocket is rather constant for low masses ($M \le 2$ \ms) while it rapidly drops to very small values for higher-mass AGB stars. The extensions of the $p$-enriched reservoir are shown for two choices of the starting layer for buoyancy: (i) that characterized by a dynamical viscosity of $\mu_d$ = 0.01 and (ii) the one where $\mu_d$ = 0.05. This second  value seems so far to be the one giving a more coherent interpretation of solar abundances \citep{tri16} and of isotopic anomalies in presolar grains \citep{pal17b}. We shall therefore adopt it as a reference here.

The same basic mechanism drives DM in the H-rich layers below the convective envelope during H-shell burning. This mixes to the envelope, with the same process, products of proton captures, including the nucleus \al, whose prediction is needed for our SLR calculations. Details on the determination of the $^{26}$Al abundance in the envelope have been already published and can be found in \citet{pal17a}.

On the basis of the above procedures, we have computed the envelope abundances of \alb and of neutron-rich nuclei for several models; we shall discuss the results for a couple of typical cases among those indicated in Table 2. 

\begin{table*}[t]
\begin{center}
{
{Table 3. Mass of the $^{13}$C pocket for different models}

\vspace{0.5cm}

\begin{tabular}{c c c}
\hline
\hline
  AGB Model & $\mu_{d} = $ 0.01 & $\mu_{d} = $ 0.05   \\  
\hline
1.5 \ms, [Fe/H] = $~~0.0$  & 2.8$\cdot 10^{-3}$ \msb & 4.9$\cdot 10^{-3}$ \msb  \\
2.0 \ms, [Fe/H] = $~~0.0$  & 2.6$\cdot 10^{-3}$ \msb & 4.4$\cdot 10^{-3}$ \msb  \\
3.0 \ms, [Fe/H] = $-0.5$  & 5.4$\cdot 10^{-4}$ \msb & 1.4$\cdot 10^{-3}$ \msb  \\
5.0 \ms, [Fe/H] = $~~0.0$  & 1.5$\cdot 10^{-5}$ \msb & 0.9$\cdot 10^{-4}$ \msb  \\
 \hline
\end{tabular}
}
\end{center}

\label{tab2}
\end{table*}

One can apply the formula of equation (5) to a couple of nuclei produced by a model star, thus fixing the two parameters $d$ and $\Delta$. Then one has to verify what kind of predictions this implies for the other radioactive nuclei in pristine solids. 

For the reference nuclei we use \alb and \ca. The motivation for this choice lays in the fact that it has been ascertained \citep{dt07,vil09} that \alb cannot be produced by solar spallation, and needs to derive from a stellar source. Moreover, the correlation between \alb and \cab established by \citet{s+98} and reported here in Figure 1 suggests that the two nuclei may have the same origin. We underline that this constraint is not considered in several published scenarios among those quoted, but this appears to be a serious drawback: in fact, the mentioned correlation and the stellar origin for \alb represent important pieces of evidence and should be taken into account.

Tables 4 and 5 show the outcomes obtained by deriving the two free parameters from the mentioned nuclei (i.e. fixing $d$ and $\Delta$ so that the measurements for \alb and \cab in ESS solids are reproduced), adopting two typical models for IMSs from our calculations. They refer to a 3 \msb and a 5 \msb star, with solar metallicity. (We do not discuss here in detail results for significantly lower masses, because of their excessively long lifetimes, essentially inhibiting any chance encounter with the forming Sun). 

The main result that can be derived from even a quick glance to Tables 4 and 5 is that, when a \ctb pocket is included (even of a minimal extension, as in the case of the 5 \msb model), accounting for the lighter radioactive isotopes \alb and \cab always implies some large excesses on nuclei heavier than iron. This is a very big problem: a deficit on isotopes like \pd, \cs, or $^{182}$Hf, which are not of purely $s$-process origin, might be compensated by some inheritance of $r$-process products from Galactic evolution; but large excesses, like those shown in the tables, cannot be accomodated. 

As mentioned, the main difference between our models and those by \citet{w+06}, where a nice solution could be found for several SLRs, is the presence of a \ctb pocket, which was instead excluded in that solution. Whatever the extension of the pocket is, the neutron flux remains always too large to find any consistency with \alb production. In the light of our MHD model for the formation of the neutron source \ct, we have no way to solve this inconsistency and must admit that a solution for SLRs in the framework of our models for mixing and nucleosynthesis in AGB stars can no longer be found.
 
\begin{table*}[t!!]
\begin{center}
{
{Table 4. SLRs as predicted by a 3 \msb model with MHD mixing}

\vspace{0.2cm}
{[Fe/H] = 0 - Dilution $d$ = 9.05$\cdot$10$^{-3}$ $-$ Delay time $\Delta$ = 0.87 Myr}

\vspace{0.2cm}
\begin{tabular}{c c c c c c c}
\hline
\hline
  Rad. & Ref. & $\tau_R$ (Myr) &$N^R/N^S$ & $q^S$ & $\alpha^{R,S}$ & $[N^R/N^S]_{Meas.}$   \\
\hline
  $^{26}$Al & $^{27}$Al  & 1.03 & 1.34$\cdot$10$^{-2}$ & 1.004 & 5.23$\cdot$10$^{-5}$ & (5.23 $\pm$ 0.13)$\cdot$10$^{-5}$  \\
  $^{41}$Ca & $^{40}$Ca  & 0.15 &1.48$\cdot$10$^{-4}$ &0.994 &4.00$\cdot$10$^{-9}$ &4$\cdot$10$^{-9}$   \\
  $^{60}$Fe & $^{56}$Fe  & 3.75 &2.18$\cdot$10$^{-5}$ &0.994 &1.55$\cdot$10$^{-7}$ & 10$^{-8} - 10^{-6}$ \\
  $^{107}$Pd & $^{108}$Pd & 9.4 & 1.33$\cdot$10$^{-1}$ & 6.198 & 6.78$\cdot$10$^{-3}$& (5.9 $\pm$ 2.2)$\cdot$10$^{-5}$   \\
  $^{135}$Cs & $^{133}$Cs  & 3.3&6.77$\cdot$10$^{-1}$ &2.101 &5.53$\cdot$10$^{-3}$ &  4.8$\cdot$10$^{-4}$\\
  $^{182}$Hf & $^{180}$Hf  &12.8&1.18$\cdot$10$^{-2}$ &4.027 &4.01$\cdot$10$^{-4}$ & (9.81 $\pm$ 0.41)$\cdot$10$^{-5}$ \\
  $^{205}$Pb &  $^{204}$Pb & 22 &6.58$\cdot$10$^{-1}$ & 2.552 &1.46$\cdot$10$^{-2}$ & 10$^{-3}$\\
\hline
\end{tabular}
}
\end{center}
\label{tab3}
\end{table*}

\begin{table*}[t!!]
\begin{center}
{
{Table 5. SLRs as predicted by a 5 \msb model with MHD mixing}

\vspace{0.2cm}
{[Fe/H] = 0 - Dilution $d$ = 3.27$\cdot$10$^{-2}$ $-$ Delay time $\Delta$ = 0.85 Myr}

\vspace{0.2cm}
\begin{tabular}{c c c c c c c}
\hline
\hline
  Rad. & Ref. & $\tau_R$ (Myr) &$N^R/N^S$ & $q^S$ & $\alpha^{R,S}$ & $[N^R/N^S]_{Meas.}$   \\
\hline
  $^{26}$Al & $^{27}$Al  & 1.03 & 3.65$\cdot$10$^{-3}$ & 1.002 & 5.23$\cdot$10$^{-5}$ & (5.23 $\pm$ 0.13)$\cdot$10$^{-5}$  \\
  $^{41}$Ca & $^{40}$Ca  & 0.15 &3.57$\cdot$10$^{-5}$ &0.996 &4.00$\cdot$10$^{-9}$ &4$\cdot$10$^{-9}$   \\
  $^{60}$Fe & $^{56}$Fe  & 3.75 &4.36$\cdot$10$^{-4}$ &0.995 &1.13$\cdot$10$^{-5}$ & 10$^{-8} - 10^{-6}$ \\
  $^{107}$Pd & $^{108}$Pd & 9.4 & 2.24$\cdot$10$^{-2}$ & 1.139 & 7.61$\cdot$10$^{-4}$& (5.9 $\pm$ 2.2)$\cdot$10$^{-5}$   \\
  $^{135}$Cs & $^{133}$Cs  & 3.3&3.09$\cdot$10$^{-2}$ &1.011 & 7.60$\cdot$10$^{-4}$ &  4.8$\cdot$10$^{-4}$\\
  $^{182}$Hf & $^{180}$Hf  &12.8&3.66$\cdot$10$^{-4}$ &1.026 &1.15$\cdot$10$^{-5}$ & (9.81 $\pm$ 0.41)$\cdot$10$^{-5}$ \\
  $^{205}$Pb &  $^{204}$Pb & 22 & 4.42$\cdot$10$^{-2}$ &1.038 &1.44$\cdot$10$^{-3}$ & 10$^{-3}$\\
\hline
\end{tabular}
}
\end{center}

\label{tab4}
\end{table*}

\subsection{Effects of a Late AGB Star. II. Models with Opacity-Induced Mixing}

\begin{figure*}[t!!]                                           
\begin{centering}
\includegraphics[width=1.0\columnwidth]{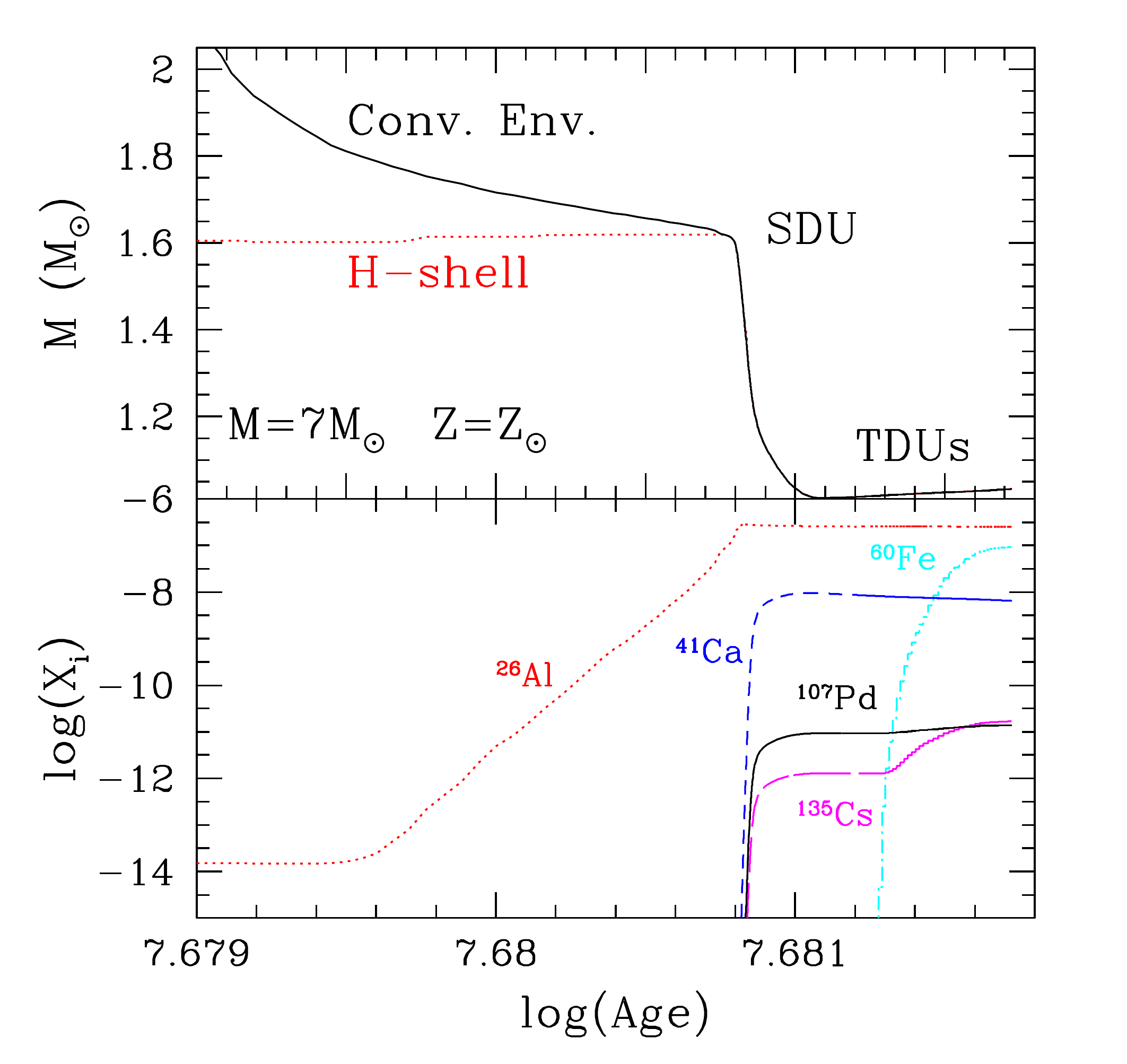}                
\caption{Upper panel: temporal evolution of the mass coordinates of the H-burning shell and of the inner border of the convective envelope for a 7 M$_\odot$ model with solar metallicity.
Lower panel: temporal evolution of surface abundances for some radioactive isotopes of interest.}
\end{centering}
\end{figure*}                                                    

During a TDU, more than one physical mechanism can contribute to the downward diffusion of protons into the underlying He- and C-rich layer. Spectroscopic observations and 
pre-solar SiC grain measurements provide indications on the shape and extension of the $^{13}$C pocket. Recent studies point to a flat $^{13}$C profile in a region from  one to a few $10^{-3}$ M$_\odot$ thick \citep[see e.g.][]{liu14,liu15,pal17b}; in this respect both the DM models considered in this work fulfil the basic requirements. Among the constraints to respect, there is also a need to limit the abundance by mass of \ctb in each layer, otherwise a too large production of $^{14}$N (a neutron poison) would be 
obtained, hampering $s$-processing. In the so-called \oq FRUITY\cqb models \citep{cri11,cri15b,cri16}, the approach first discussed by \citet{beckibe} was followed. It considers the effects on mixing of the opacity difference between the (opaque) H-rich envelope and the (more transparent) He-rich underlying region. During a TDU, such a difference produces a discontinuity in the temperature gradient at the inner border of the envelope, which leads to a consequent abrupt decrease of the convective velocities (see \citealt{stra06}). This process is unbalanced and unstable: any perturbation of the convective/radiative interface tending to expand downward the boundary would grow, thus leading to an even deeper mixing. In such a condition, it is reasonable to hypothesize that the result is not an extension of the envelope as a whole, but rather that individual convective elements with non-zero velocities penetrate beyond the limit defined by the Schwarzschild criterion (where the radiative gradient equals the adiabatic one). Those bubbles are decelerated by the steep pressure gradient immediately below the inner envelope border, which strongly  limits the extent of their penetration. In order to mimic this behavior, 
we imposed that bubble velocities below the formal Schwarzschild border  decline exponentially, namely:

\begin{equation} \label{param}
v=v_{IN}\exp{
\left(
-\frac{\Delta r}{\beta H_p}
\right)
} \;     
\end{equation}

Here $\Delta$r is the distance from the Schwarzschild border, $v_{IN}$ is the velocity of the most internal convective mesh, $H_p$ is the pressure scale height at the border itself and $\beta$ is a free parameter \citep[usually set to $\beta$=0.1 on observational grounds: for its calibration see][]{cri09}. The introduction of this algorithm implies that:

\begin{enumerate}
\item the convective border becomes more stable;
\item the TDU efficiency is increased;
\item a profile of protons is left below the convective envelope.
\end{enumerate}

As with MHD instabilities discussed in the previous Subsection, also in this approach the mass extension of the pocket does not remain  constant along the AGB, but decreases steadily, following the shrinking of the He-intershell region with increasing core mass \citep{cri09}. A potential problem arising from this approach is that the exponential decline of convective velocities would proceed to the center of the star, unless a maximum penetration is fixed. This limit was initially set to 2$H_p$ \citep{stra06}. Later, a better match to isotopic ratios in pre-solar SiC grains \citep{liu14,liu15} suggested to fix the penetration limit in terms of the convective velocity, imposing that it stops at a certain small fraction ($10^{-11}$) of the value achieved at the Schwarzschild border. This corresponds to a depth  $2.2 - 2.3 H_p$. 

The $^{13}$C pockets obtained with the method outlined above are not remarkably different from those proposed by \citet{tri16} and discussed in Subsection 5.1, although they are characterized by a larger amount of $^{14}$N in the upper region. This feature is intrinsically connected to the approach followed, which yields a top-down flow of the material (and not a bottom-up movement, like for the magnetic tubes characterizing the models of Subsection 5.1). 

We refer to \citet{cri15a,cri16} and references therein for a detailed description of the nucleosynthesis resulting from the assumptions outlined above. Here we recall simply that the models discussed are rather \oq cool\cqb (e.g. cooler than those outlined in Subsection 5.1) and due to this do not experience HBB, a process that, as already mentioned, has a strong model dependence. 

For the specific purposes of this work and in the framework of the approach just discussed, we integrated the FRUITY database by computing also more massive AGB models (6.5 $\le$ M/\msb $\le$ 8) at solar metallicity. The surface isotopic distributions of those masses include the effects of SDU. This event has a considerable impact on radioactive isotope abundances. As already highlighted, SDU occurs during the early-AGB phase of IMSs; at that time, the switching off of the H-burning shell is followed by the inward penetration of the convective envelope in the H-depleted zone (upper panel of Figure 3). 
As widely reported in the literature, among the consequences of the SDU there is an increase of the surface $^{4}$He, $^{14}$N, and $^{26}$Al abundances. A further effect present in our models induces enhancements for at least three more SLRs of interest, beyond \al: they are $^{41}$Ca, $^{107}$Pd, and $^{135}$Cs, ensuing from a marginal neutron-capture episode (see the lower panel of Figure 3, which also shows \fe).  Indeed, those isotopes are produced by the radiative burning of the amount of $^{13}$C present in H-burning ashes (its CNO equilibrium abundance). This occurs because the layers beyond the H-burning shell are heated up to more than $10^8$ K, i.e. to large enough temperatures to activate the $^{13}$C($\alpha$,n)$^{16}$O reaction. Later, surface abundances are further changed by TDUs (although not for all isotopes). Note that a similar finding has never been reported in the literature, because post-process calculations commonly ignore the nucleosynthesis of heavy elements before the TP-AGB phase. Our FRUITY stellar evolutionary models, instead, are computed with a full nuclear network, starting from the Main Sequence phase and up to the tip of the AGB.

Once all the above effects are considered, the final surface abundances of radioactive nuclei in our  models can be used to estimate their possible contribution to their inventory in the ESS. Tables 6 and 7 summarize these results for two typical cases. Much like what we obtained in Subsection 5.1, also in this case there is no space for a compromise agreement. The predictions always include  excesses of some heavy (A $>$ 56) neutron-capture nuclei (especially \pd) with respect to \alb and \cab (from which, again, we deduce the time delay $\Delta$ and the dilution factor, $d$). We argue that this conclusion is not limited to the specific models considered in  this work as examples: any DM model yielding proton penetration into the He-rich layers, inducing the formation of a \ctb pocket, will inevitably end up with excesses of $^{107}$Pd and other neutron-rich isotopes with respect to the lighter ones.

\begin{table*}[t!!]
\begin{center}
{
{Table 6. SLRs as predicted by a 6 \msb model.}

\vspace{0.2cm}
{[Fe/H] = 0 - Dilution $d$ = 5.18$\cdot$10$^{-2}$ $-$ Delay time $\Delta$ = 0.98 Myr}

\vspace{0.2cm}
\begin{tabular}{c c c c c c c}
\hline
\hline
  Rad. & Ref. & $\tau_R$ (Myr) &$N^R/N^S$ & $q^S$ & $\alpha^{R,S}$ & $[N^R/N^S]_{Meas.}$   \\
\hline
  $^{26}$Al & $^{27}$Al  & 1.03 & 2.64$\cdot$10$^{-3}$ & 0.991 & 5.23$\cdot$10$^{-5}$ & (5.23 $\pm$ 0.13)$\cdot$10$^{-5}$  \\
  $^{41}$Ca & $^{40}$Ca  & 0.15 &5.42$\cdot$10$^{-5}$ &0.980 &4.00$\cdot$10$^{-9}$ &4$\cdot$10$^{-9}$   \\
  $^{60}$Fe & $^{56}$Fe  & 3.75 &1.32$\cdot$10$^{-5}$ &1.008 &5.30$\cdot$10$^{-7}$ & 10$^{-8} - 10^{-6}$ \\
  $^{107}$Pd & $^{108}$Pd & 9.4 & 2.31$\cdot$10$^{-2}$ & 1.200 & 1.29$\cdot$10$^{-3}$& (5.9 $\pm$ 2.2)$\cdot$10$^{-5}$   \\
  $^{135}$Cs & $^{133}$Cs  & 3.3&3.75$\cdot$10$^{-2}$ &1.007 & 1.39$\cdot$10$^{-3}$ &  4.8$\cdot$10$^{-4}$\\
  $^{182}$Hf & $^{180}$Hf  &12.8&1.21$\cdot$10$^{-2}$ &1.271 &7.34$\cdot$10$^{-4}$ & (9.81 $\pm$ 0.41)$\cdot$10$^{-5}$ \\
  $^{205}$Pb &  $^{204}$Pb & 22 & 1.73$\cdot$10$^{-2}$ &1.099 &9.39$\cdot$10$^{-4}$ &10$^{-3}$\\
\hline
\end{tabular}
}
\end{center}

\label{tab5}
\end{table*}

\begin{table*}[t!!]
\begin{center}
{
{Table 7. SLRs as predicted by a 7 \msb model.}

\vspace{0.2cm}
{[Fe/H] = 0 - Dilution $d$ = 3.32$\cdot$10$^{-2}$ $-$ Delay time $\Delta$ = 1.01 Myr}

\vspace{0.2cm}
\begin{tabular}{c c c c c c c}
\hline
\hline
  Rad. & Ref. & $\tau_R$  (Myr) &$N^R/N^S$ & $q^S$ & $\alpha^{R,S}$ & $[N^R/N^S]_{Meas.}$   \\
\hline
  $^{26}$Al & $^{27}$Al  & 1.03 & 4.31$\cdot$10$^{-3}$ & 0.976 & 5.23$\cdot$10$^{-5}$ & (5.23 $\pm$ 0.13)$\cdot$10$^{-5}$  \\
  $^{41}$Ca & $^{40}$Ca  & 0.15 &1.05$\cdot$10$^{-4}$ &0.980 &4.00$\cdot$10$^{-9}$ &4$\cdot$10$^{-9}$   \\
  $^{60}$Fe & $^{56}$Fe  & 3.75 &6.79$\cdot$10$^{-5}$ &1.008 &1.74$\cdot$10$^{-6}$ & 10$^{-8} - 10^{-6}$ \\
  $^{107}$Pd & $^{108}$Pd & 9.4 & 1.21$\cdot$10$^{-2}$ & 1.078 & 3.88$\cdot$10$^{-4}$& (5.9 $\pm$ 2.2)$\cdot$10$^{-5}$   \\
  $^{135}$Cs & $^{133}$Cs  & 3.3&1.24$\cdot$10$^{-2}$ &0.963 & 2.79$\cdot$10$^{-4}$ &  4.8$\cdot$10$^{-4}$\\
  $^{182}$Hf & $^{180}$Hf  &12.8&7.24$\cdot$10$^{-3}$ &1.132 &2.51$\cdot$10$^{-4}$ & (9.81 $\pm$ 0.41)$\cdot$10$^{-5}$ \\
  $^{205}$Pb &  $^{204}$Pb & 22 & 1.03$\cdot$10$^{-2}$ &0.992 &3.23$\cdot$10$^{-4}$ &10$^{-3}$\\
\hline
\end{tabular}
}
\end{center}

\label{tab6}
\end{table*}

Figure 4 summarizes the results of this Section, showing together the predictions from Subsections 4.1 and 4.2; they can be represented together thanks to the similarity of the findings, despite the different (complementary) starting hypotheses. Over the mass range from about 2 to about 8 \ms, when one constrains the free parameters $\Delta$ and $d$ through a fit of the measured ESS abundances of \alb and \ca, heavier nuclei are not simultaneously accounted for and in all cases the most relevant problem resides in a huge overproduction of \pd. This fact leads us to the conclusion that the scenario of the Solar System pollution by an AGB star of intermediate mass may be inadequate, although for the most massive models 
a window is still open. In particular, our models are rather cool and do not develop HBB. As mentioned, this process was examined by \citet{was17}, with negative conclusions. However, HBB is still strongly model dependent. One would need an ad-hoc H-burning process at the base of the envelope, to produce \alb in exactly the right amount to compensate for the excesses in $n$-rich isotopes induced by the presence of a \ctb pocket. The reality of this possibility must however be explored in detail. This will be the subject of a forthcoming work.

\begin{figure*}[t!!]                                               
\begin{centering}
\includegraphics[width=0.8\columnwidth]{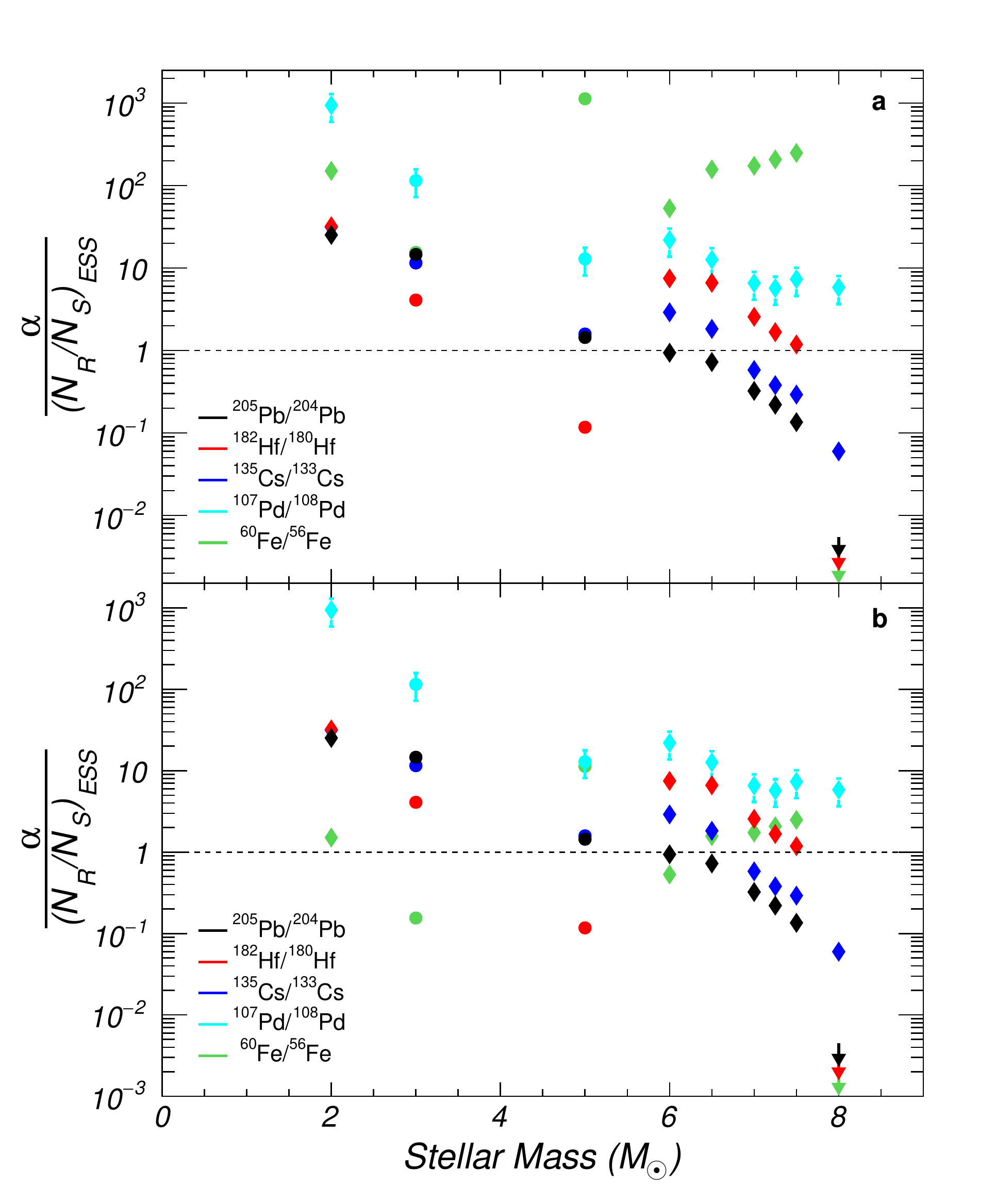}                
\caption{Prediction for heavy SLRs from AGB stars with the physical models for DM discussed in Subsections 4.1 and 4.2. In panel a) we adopt as a reference ESS ratio for \fe/$^{56}$Fe the value 10$^{-8}$, while in panel b) the opposite extreme choice is made (10$^{-6}$). The picture shows that, when the free parameters $\Delta$ and $d$ are fixed through a fit of the measured ESS abundances of \alb and \ca, heavier nuclei are not satisfactorily accounted for. Just to make an example, even the models explainig well \feb (those around 6$-$7 \ms, in panel b) always imply excesses by a factor $8 - 20$ on $^{107}$Pd. (In the plot circular dots indicate the models of Subsection 5.1, diamonds those of Subsection 5.2. The case of a 2 \msb illustrated here is not discussed explicitly in a table, due to its too long lifetime that would make  a chance encounter with the forming Sun impossible. It is shown here only to illustrate the rather smooth and continuous sequence of behaviors characterizing the two series of models that, albeit different, converge to a rather unique and coherent view of AGB phases.)}
\end{centering}
\vspace{0.2cm}
\end{figure*}   

\section{Nucleosynthesis in Massive Stars and the Role of Late Supernovae}
\label{sec:MS}

\begin{figure}[t!!]                                              
\begin{centering}
\includegraphics[width=0.6\columnwidth]{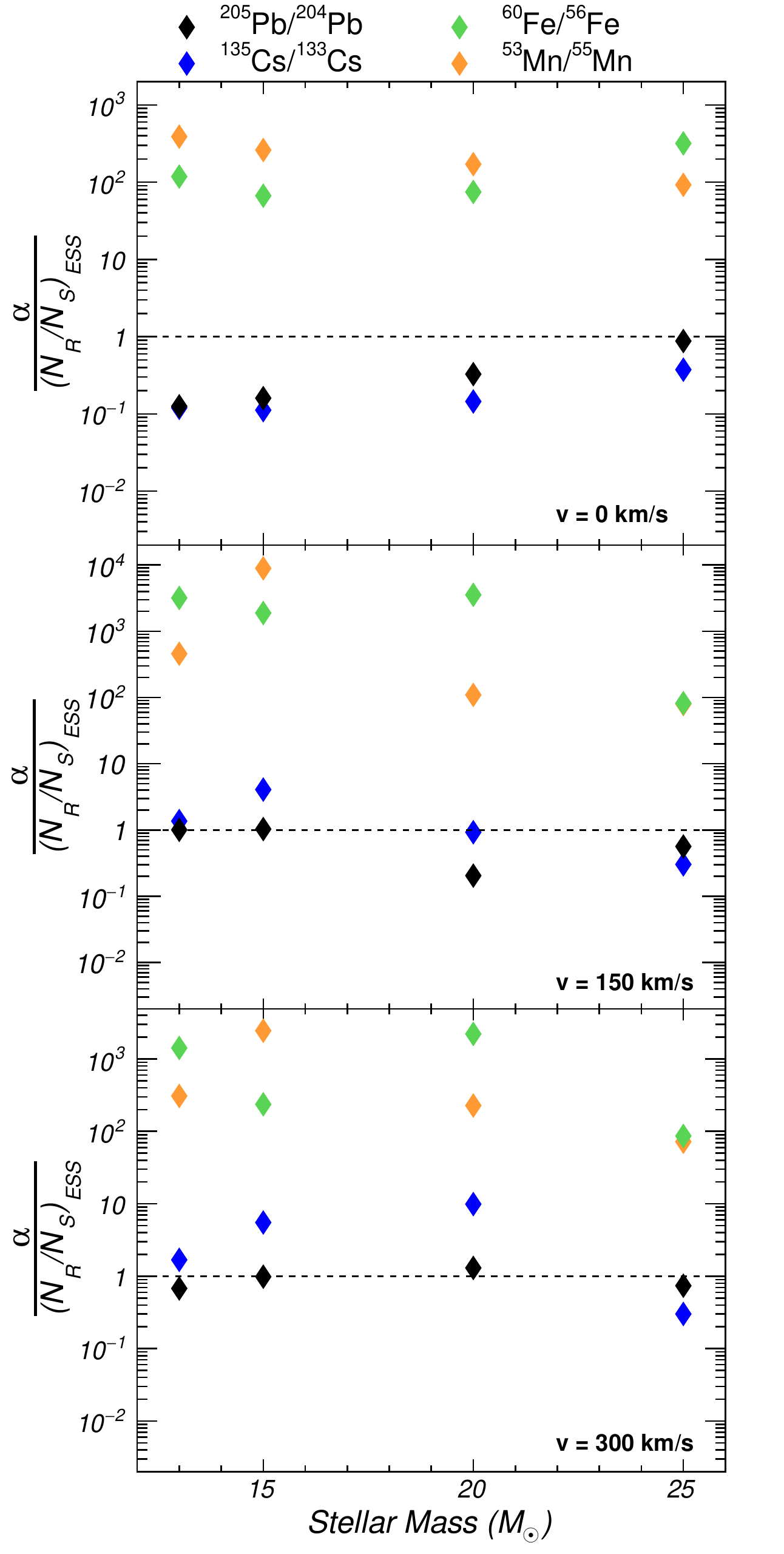}                
\caption{Predictions for SLRs from a late contamination by a CCSNe in the mass range 13$-$25 \ms, assuming as a reference a solar \fe/$^{56}$Fe ratio of 1$\cdot 10^{-8}$. No one of the models shown can account acceptably for the measurements, once \alb and \cab are used for fixing the free parameters. In particular, \feb and $^{53}$Mn would be in any case enormously overproduced.}
\end{centering}
\end{figure}  
 
\begin{figure}[t!!]                                              
\begin{centering}
\includegraphics[width=0.58\columnwidth]{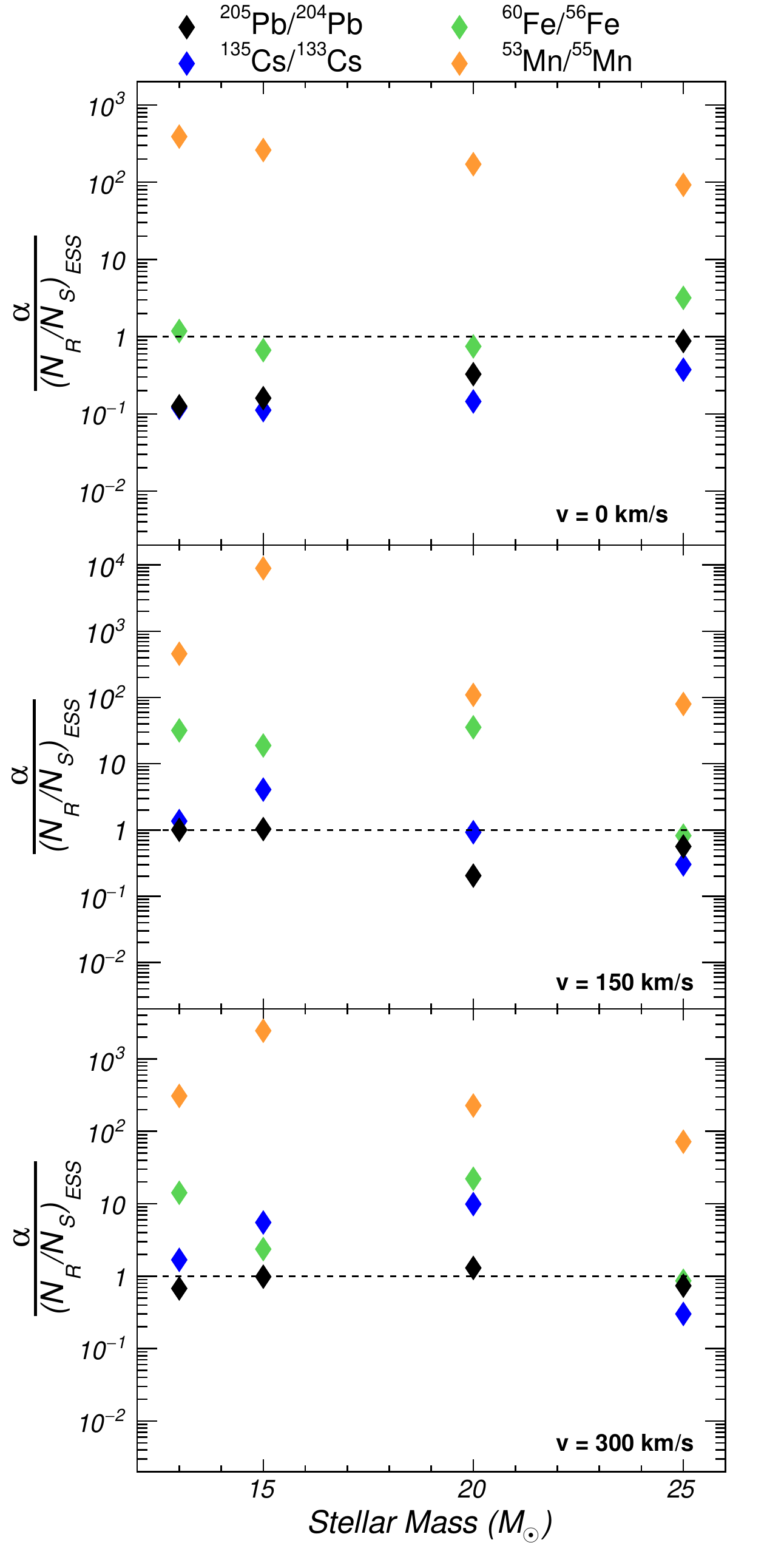}                
\caption{Predictions for SLRs from a late contamination by a CCSN in the mass range 13-25 \msb assuming as a reference a solar \fe/$^{56}$Fe ratio of 1$\cdot 10^{-6}$. As the figure shows, in this second case the models with no rotation up to 20 \msb and  the model of a 25 \msb with a rather large rotation rate ($\geq$ 150 km/sec) would account reasonably for \feb (in addition to \alb and \cab that were used for fixing the parameters) avoiding overproductions for $^{107}$Pd and $^{135}$Cs. However, $^{53}$Mn would remain, also in this case, overproduced by two orders of magnitude and would require a much larger mass cut, as in \citet{mc00}.}
\end{centering}
\end{figure}

\begin{figure}[t!!]
\begin{centering}                                              
\includegraphics[width=0.6\columnwidth]{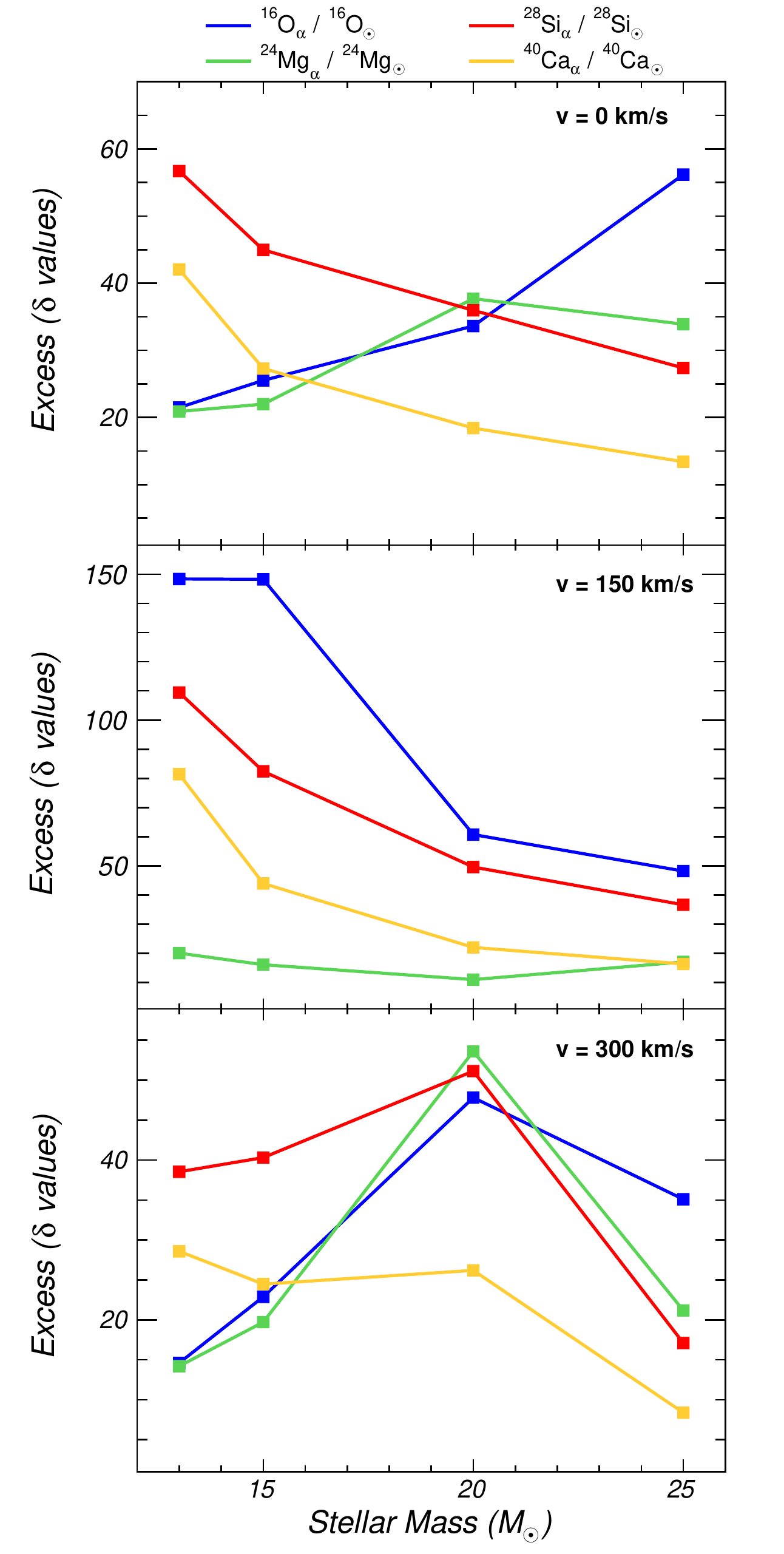}                
\caption{Values of the excesses introduced on $\alpha$-rich isotopes of major stable 
elements by the CCSN models discussed so far, tuned to account for the \al/$^{27}$Al and \ca/$^{40}$Ca number ratios in the ESS. The plot shows the increase with respect to accepted average abundances. We recall that shifts in excess of a few permil are excluded by present-day measurements (see Section 2).}
\end{centering}
\end{figure}                                                 

MSs are defined as those that can contract and heat up to several billions degrees, eventually reaching the conditions for the collapse of the core up to nuclear densities. Depending on the initial metallicity, the initial rotational velocity and the treatment of the convective borders, the minimum mass suitable to evolve up to the final stages ranges somewhere between 9 and 12 \ms. This limit marks also the maximum mass of the so-called Super-AGB stars, i.e. those that form a strongly electron-degenerate core only after burning C and/or Ne. The evolution of these stars is characterized by repeated, weak but very frequent, thermal pulses and may end up with the formation of O-Ne white dwarfs or electron-capture supernovae \citep{den13,dohe}.  The possible contributions of Super-AGB stars to the inventory of ESS radioactivities was early analyzed by \citet{lug12}; it will not be re-discussed here, in view of the large uncertainties and model-dependencies related to the very complex physics still affecting this mass range.

Since massive stars avoid electron degeneracy and evolve towards higher and higher temperatures, they activate nuclear reactions that form increasingly heavier nuclei through the H-, He-, C-, Ne-, O-, and Si-burning phases. Once the temperature reaches $\sim~10$ GK, electrons become relativistic and the contraction reverses into a collapse of the core. Here  nuclear densities are reached, eventually driving the phenomenon of a CCSN explosion. The ensuing shock wave causes the violent expulsion of the external layers into the interstellar medium. The very high peak temperature achieved by the innermost layers, while the shock front moves outward, induces a substantial modification of the pre-existing composition; this means that it is not correct to neglect the nucleosynthesis induced by the passage of the shock wave (explosive nucleosynthesis) if one wants to study the contribution of massive stars to the Galactic enrichment and to the ESS composition. Unfortunately, some of the works present in the literature on the ESS radioactivities do not take into account explosive phases. This is in particular the case for some of the scenarios for the sequential contamination of a pre-solar molecular cloud \citep{gm12,gou06}. One should therefore look at these results with a bit of caution. 

\begin{table*}[t!!]
\begin{center}
{Table 8. SLRs as predicted by a non-rotating 20 \msb model.}

\vspace{0.2cm}
{[Fe/H] = 0 - Dilution $d$ = 1.85$\cdot$10$^{-4}$ $-$ Delay time $\Delta$ = 1.34 Myr}

\vspace{0.2cm}
\begin{tabular}{c c c c c c c}
\hline
\hline
  Rad. & Ref. & $\tau_R$ (Myr) &$N^R/N^S$ & $q^S$ & $\alpha^{R,S}$ & $(N^R/N^S)_{Meas.}$   \\
\hline
  $^{26}$Al & $^{27}$Al  & 1.03 & 5.42$\cdot$10$^{-3}$ & 192.2 & 5.23$\cdot$10$^{-5}$ & (5.23 $\pm$ 0.13)$\cdot$10$^{-5}$  \\
  $^{41}$Ca & $^{40}$Ca  & 0.15 & 1.68$\cdot$10$^{-3}$ &99.52 &4.00$\cdot$10$^{-9}$ &4$\cdot$10$^{-9}$   \\
  $^{53}$Mn & $^{55}$Mn  & 5.3 & 5.75$\cdot$10$^{-1}$ &13.88 &1.15$\cdot$10$^{-3}$ &(6.7 $\pm$ 0.56)$\cdot$10$^{-6}$  \\  
  $^{60}$Fe & $^{56}$Fe  & 3.75 &3.44$\cdot$10$^{-4}$ &16.80 &7.48$\cdot$10$^{-7}$ & 10$^{-8} - 10^{-6}$ \\
  $^{135}$Cs & $^{133}$Cs  & 3.3&4.13$\cdot$10$^{-2}$ &14.43 &6.95$\cdot$10$^{-5}$ &  4.8$\cdot$10$^{-4}$\\
  $^{205}$Pb &  $^{204}$Pb & 22 &1.54$\cdot$10$^{-1}$ & 12.19 &3.27$\cdot$10$^{-4}$ & 10$^{-3}$\\
\hline
\end{tabular}
\end{center}
\label{tab7}
\end{table*}

\begin{table*}[t!!]
\begin{center}
{Table 9. SLRs as predicted by a 25 \msb model with rotational velocity of 150km/s.}

\vspace{0.2cm}
{[Fe/H] = 0 - Dilution $d$ = 9.11$\cdot$10$^{-5}$ $-$ Delay time $\Delta$ = 1.41 Myr}

\vspace{0.2cm}
\begin{tabular}{c c c c c c c}
\hline
\hline
  Rad. & Ref. & $\tau_R$ (Myr) &$N^R/N^S$& $q^S$ & $\alpha^{R,S}$ & $(N^R/N^S)_{Meas.}$   \\
\hline
  $^{26}$Al & $^{27}$Al  & 1.03 & 1.08$\cdot$10$^{-2}$ & 209.5 & 5.23$\cdot$10$^{-5}$ & (5.23 $\pm$ 0.13)$\cdot$10$^{-5}$  \\
  $^{41}$Ca & $^{40}$Ca  & 0.15 & 3.01$\cdot$10$^{-3}$ &179.7 &4.00$\cdot$10$^{-9}$ &4$\cdot$10$^{-9}$   \\
  $^{53}$Mn & $^{55}$Mn  & 5.3 & 5.44$\cdot$10$^{-1}$ &14.08 &5.35$\cdot$10$^{-4}$ &(6.7 $\pm$ 0.56)$\cdot$10$^{-6}$  \\  
  $^{60}$Fe & $^{56}$Fe  & 3.75 &7.38$\cdot$10$^{-4}$ &17.80 &8.21$\cdot$10$^{-7}$ & 10$^{-8} - 10^{-6}$ \\
  $^{135}$Cs & $^{133}$Cs  & 3.3&1.43$\cdot$10$^{-1}$ &18.10 &1.45$\cdot$10$^{-4}$ &  4.8$\cdot$10$^{-4}$\\
  $^{205}$Pb &  $^{204}$Pb & 22 &3.05$\cdot$10$^{-1}$ & 21.63 &5.64$\cdot$10$^{-4}$ & 10$^{-3}$\\
\hline
\end{tabular}
\end{center}

\label{tab8}
\end{table*}

In the external layers of the star, through which the shock wave passes, the relics of core-He- and shell-C-burning are abundant. These zones include \textit{s}-process isotopes, having experienced \textit{n}-captures, mainly from the \neanb source, during the quiet evolution of the supernova progenitor. They are actually considered as the main site where the $weak$ $s$-process component is generated \citep{rai,pig}, producing isotopes up to those of Sr, sited at the \oq magic\cqb neutron number N=50. Above this mass region, neutron captures in massive stars become rather inefficient, but they might still play a role for the ESS radioactivities, as these last were typically produced in environments with small enhancement factors $q^s$(see also Section 5).

In the scenario of Solar System formation, a massive star forming a CCSN was suggested by several authors as a potential trigger, for inducing the final collapse of the presolar cloud through shock waves  \citep{ct77,vb02,boss}. This event must have been accompanied by a strong injection of newly formed nuclei into the ESS, including SLRs like \al, \ca, \fe, $^{53}$Mn, plus possibly those synthesized in fast and slow $n$-capture processes. Also many stable isotopes of major elements are expected to receive important contributions. Hence, by modelling the processes of hydrostatic and explosive nucleosynthesis in massive pre-supernova and during the final CCSN events, one can predict the isotopic anomalies that such a pollution would have induced in the solar nebula. 

As already mentioned, CCSNe are typically the main producers of intermediate-mass elements, from O to Ti; they also synthesize C, iron-peak nuclei and the weak component of the \textit{s}-process. In order to evaluate the possibility that a late Supernova event be responsible for polluting the ESS with the radioactive species not accounted for by the average Galactic enrichment (Section 3),  it is therefore also necessary to estimate the total variations that this would imply both on the solar abundances of stable elements and on their isotopic admixture. These contributions strongly depend on the initial stellar mass, on the chemical composition and on the rotational properties of the stellar models. 

In this work we consider, as possible contaminating candidates, stars that are parents to CCSNe, having a solar metallicity and an initial mass in the range 
from 13 to 25 M$_{\odot}$. The models were computed  with the FRANEC\footnote{\textit{Software:} Frascati RAphson-Newton Evolutionary Code \citep{cl13,cl15,cl18}.} evolutionary code \citep{cl13,cl15,cl18} and include the effects of rotation. The main consequence of including rotation is that of increasing the total yields of the elements, because 
it basically produces additional mixing processes, 
thus feeding more efficiently the burning layers with
fresh fuels. For our purposes, we consider a set of
models with masses of 13, 15, 20 and 25 M$_{\odot}$ 
and initial equatorial rotational velocities of 0, 150
and 300 km/s. Each model was  evolved from the pre-main
sequence up to the onset of the core bounce with the FRANEC code. The explosive phases were then computed by re-processing the structure and composition left after
the hydrostatic phases through a hydrodynamical code,
which takes into account the passage of the shock wave
and the occurrence of explosive nucleosynthesis. The
explosion was simulated in each case by considering 
the mixing and fall back mechanism \citep{UN02}. 
Within the mixing and fall back scenario, it is 
assumed that, after the passage of the shock wave, a
fraction of the most internal zone of the star is
homogeneously mixed before the fallback on the remnant
of part of the ejected material occurs. In these models the inner border of the mixed region was set at the layer where [Ni/Fe]=0.2, while the outer border was fixed at the base of the O shell (X(O)=0.001). Then, the mass cut between ejected and non-ejected material is chosen by requiring that 0.07 M$_{\odot}$ of $^{56}$Ni are ejected, thus reproducing the known iron production from SN1987A \citep{li+93}.

Tables 8 and 9 show how all the models considered do produce SLRs, especially $^{53}$Mn and \fe, In particular, fixing again the dilution factor $d$ and the time delay $\Delta$ so that \alb and \cab are reproduced, we see that the 25 \msb model, in cases characterized by a fast rotation velocity and adopting the highest choices of the \feb/$^{56}$Fe ratio in the ESS (close to $\simeq$ 10$^{-6}$), would account rather well for \feb itself and would imply a ratio $^{205}$Pb/$^{204}$Pb within a factor-of-two from the measurement. It would however also yield some deficit in $^{135}$Cs, which, as seen, cannot be compensated by contributions from the Galactic evolution of $r$-process nuclei (despite the fact that this isotope is not shielded against fast $r$-process decays). $^{53}$Mn would be overabundant by almost two orders of magnitude and would require a special mass cut, as suggested by \citet{mc00}. ($^{107}$Pd and $^{182}$Hf were not in the network adopted in the original models and their abundances cannot be checked).

These results are reported also in Figures 5 and 6, for the two extreme choices of the \fe/$^{56}$Fe abundance in the ESS (10$^{-8}$ for Figure 5 and 10$^{-6}$ for Figure 6). As is shown, only in the second case, for a 25 \msb model with a high rotation rate ($\geqslant$ 150 km/sec) one obtains a reasonable agreement between some of the measurements and predictions. However, $^{53}$Mn remains largely overproduced (by two orders of magnitude). In any case, the scenario of Figure 6 requires a very large \feb abundance in the ESS. 

Furthermore, one has to notice that, even fixing ad-hoc the mass cut for accommodating $^{53}$Mn, unsolved problems would remain for the abundances of stable isotopes. This is so to the point that we have serious doubts that anyone of the cases studied can be reconciled with the measurements. This includes the models mentioned above, fitting most of the SLRs except for $^{53}$Mn. We note that the exercise of adding these ejecta to the forming star in a very simple way, with complete and homogeneous mixing of the two components, would simply shift abundances of stable nuclei without producing isotopic heterogeneities. However, more realistic scenarios that consider the possible clumpiness of ejecta or mass segregation imply the introduction of shifts on stable isotope abundances at levels incompatible with the limits set by actual measurements, as discussed previously. Just to make an example, let's assume that these shifts be of the same order of magnitude 
of the average variations introduced on stable elements in the mentioned simple exercise.  These last are shown in Figure 7 for the isotopes $^{16}$O, $^{24}$Mg, $^{28}$Si and $^{40}$Ca in term of permil (delta) values. As shown in the plots, the anomalies predicted are at least a few percent. This is much larger than allowed by present uncertainties in the meteoritic data, as discussed in Section 2. The fact that the injection process is complex and involves either the formation of clumpy structures or the incorporation of only part of the material ejected was addressed by various authors \citep{pan, gw00, mae} and seems to be required by the same inhomogeneities of observed SN remnants. Clearly, hydrodynamical models of these mixing processes should be performed considering all the SLRs, to complement the cases recently studied by \citet{dw+18} for \alb and \fe.

In view of these complexities, and of the fact that very high shifts are found in Figure 7 for elements produced in widely different layers of the star, very ad-hoc hypotheses seem to be required by any mixing model aiming at eliminating them. We also recall that effects of the same order as found by us were present in the specific model of a 11.8 \msb star by \citet{b+16}. Those authors considered the excesses found on stable isotopes as acceptable; but some of them are well in excess of 1\% so that they meet the same problems encountered here. 

A crucial and subtle problem concerning the anomalies on stable nuclei introduced by the mixing of freshly added material from an exotic source was underlined years ago by \citet{nic}. Computing such stable shifts involves mixing model yields with measured abundances. Systematic errors in the model yields can give unrealistic estimates of stable isotope anomalies. \citet{nic} tried to address this issue by mixing stellar ejecta into proxy compositions derived from chemical evolution calculations. These last used the same stellar yields also adopted in the injected matter, to normalize out the errors. According to these calculations by \citet{nic} expected anomalies in stable isotope abundances are typically at the permil level with only few outliers at the percent level. Overall, inferred anomalies are lower than predicted here (see Fig. 7) but are in many cases also incompatible with the meteorite data (see Section 2).

\section{Conclusions}
\label{sec:concl}
In this paper we have discussed a series of problems emerging in the attempts of accounting for the presence of isotopic anomalies in early solids of the solar system, induced by the {\it in situ} decay of SLRs. Such problems become really difficult to handle if one wants to get a comprehensive scenario, indicating a series of Galactic processes capable to account for nuclides with lifetimes in the range of 10-20 Myr, like $^{107}$Pd, $^{129}$I and $^{182}$Hf, and for the shorter-lived isotopes \al, \ca, \mn, \fe, and \cs. This leads us to conclude that  a self-consistent understanding of the astrophysical origins for
all the mentioned anomalies is still far from being obtained. In order to limit the uncertainties and update previous works on the subject, for the longer-lived nuclei we have tried to outline possible scenarios emerging from present-day lively debates on the astrophysical sources of the $r$-process, tentatively identifying in NSM nucleosynthesis, or in a complex, short-term granularity of the Galactic admixture of the contributions from different sources, the origin of the low abundance of $^{129}$I relative to the other neutron-rich nuclides. We indicate the first hypothesis as the most probable today. For isotopes of lifetime lower than about 5 Myr we have instead pursued our analysis on the basis of very recent calculations of stellar evolution, from 2 to 25 \ms. For IMSs, this implies the reference to current physical models for DM processes and to their implications for the activation of the neutron source \ctan. For massive stars, the approach adopted includes up-to-date models with rotation, also accounting in detail for the final explosive phases. We believe that some of the frequently quoted discussions presented so far did not consider adequately the above issues;  when this is done properly, several embarrassing open questions do remain.

In general, we showed that the continuous nuclear evolution of the Galaxy over time scales of the order of 10 Gyr might account for the radioactivities measured in the early solar nebula having lifetimes $\tau_R$ longer than about 5 Myr, including \mn. As mentioned, it is possible that the inclusion in this picture of peculiar NSM events, yielding large ratios between the abundances of nuclei at the third and second $r$-process peak, permit also an explanation for the low abundance of $^{129}$I. Shorter-lived isotopes cannot however be explained in this way. In particular, in addition to the known problematic cases of \al, \ca, and (possibly) \fe, we showed that also \csb requires an ad-hoc, late minute contribution. There is therefore a necessity to assume a late addition of nucleosynthesis products for explaining the presence of very-short lived nuclei, including the exceptionally abundant \al, which still poses the mentioned unanswered questions.

In the case of a late massive star, two relevant problems exist, which go beyond its capability of accounting for this or that SLR in a suitable amount. The first is the 
apparently unavoidable introduction into the forming solar nebula of large anomalies on stable isotopes that are excluded by present measurements. The second difficulty, recognized since a long time, concerns the possibility that the fast winds of a SN explosion can really interact with a star-forming cloud without disrupting it and instead be homogeneously mixed with the cool material of the cloud itself. Also from the point of view of the quantitative yields in SLRs, a late massive star would need very special conditions to avoid introducing enormous excesses in $^{53}$Mn, while it might fulfil the requirements posed by \al, \ca, \csb and perhaps also \pb.

The problem of mixing fast SN winds with cool condensing matter in a star forming region might be avoided in the case, suggested recently, of a sequential pollution of the molecular cloud where the Sun was born, in which the last contribution might come from a dense shell formed at the external border of a WR wind, transporting cool matter and dust \citep{gm12, d+17,dw+18}. One might guess that these models might or might not encounter the mentioned problems of excesses on stable isotopes, depending on the dilution factor and the degree of homogeneization. However, some of the most quoted discussions
in the literature do not consider the effects of previous SNe with an adequate and detailed computation of explosive nucleosynthesis; should this be done, we believe that the problems encountered here with a single star pollution would remain. Just to make a simple example, if the Sun was born in a giant molecular cloud similar to the ones we know today \citep[having masses from 10$^5$ to 3$\times$10$^6$ \ms, see e.g.][]{solomon}, then the mixing over the whole cloud of typically 20 \msb of ejecta from a CCSN would yield a dilution factor in the range from 2$\times 10^{-4}$ (similar to that of Table 8) to 0.6$\times 10^{-5}$. In the first case, the process of homogenization of the ejecta would probably be incomplete, so that, much like in Figure 7, excesses on stable nuclei at levels incompatible with measurements would remain; in the second, the excessive dilution would not explain any SLR except perhaps \fe. Possible improvements on the above not encouraging views might come from considerations of asphericities in the SN ejecta, whose effects on the solar system formation are however at the moment only speculations \citep{d+17}.
We stress again that any attempt at obtaining a compromise solution in this field should necessarily be based on quantitative models for explosive processes and for their nucleosynthesis.

Concerning less massive stars, ending their life-cycle with the described TP-AGB phases, their models recently evolved from being fully parameterized to considering the physical mechanisms that can induce non-convective mixing. These mechanisms also control the introduction of proton flows into the He-rich layers at TDU. It is found that all stars, up to at least $7 - 8$ \ms, would form a reservoir of \ct. This is bound to induce a neutron production through the \ctanb reaction, an occurrence that looks fatal for the old scenario devised by \citet{w+94,b+03,w+06}. Indeed, the neutron fluences ensuing from it, adding their effects on those from the \neanb reaction, always yield large excesses of neutron-capture radioactive isotopes, especially \pd, with respect to \al. In order to limit model dependencies, this fact was shown to occur with reference to two different treatments for DM mechanisms.

A limited parameter space remains to be explored, in the picture of a single late nucleosynthesis episode, for attempting to explain the shortest-lived radioactivities of the ESS (those with $\tau_R \lesssim 5$ Myr). One possibility resides in the already mentioned Super-AGB stars, objects in a thin mass interval (from about 9 to about 11 \ms, variable with the metallicity) between IMSs and Massive Stars. Since the mass of the \ctb pocket steadily decreases for increasing initial mass of the parent star, in that mass region it might go essentially to zero, thus avoiding the extra production of neutron-rich nuclei that in the present work appears to hamper the possibilities previously envisaged by \citet{w+06} and by \citet{tr09}. An alternative might also be found in the most massive among IMSs ($7 - 8$ \ms) if HBB calculations, which are still largely model-dependent, were found to be less effective than discussed by \citet{was17}, thus allowing to obtain more limited \al/\pdb ratios.

Coming instead to the schemes of the sequential contamination of a pre-solar molecular cloud, one should find a path in the fine-tuning of the many free parameters involved in those models, such that the CCSNe occurred in early epochs of the molecular cloud life do not introduce too large excesses on stable isotopes and nevertheless succeed in producing in adequate quantities, the required SLRs not accounted for by a possible last (and non-exploding) WR event (like e.g. \ca, \mn, \cs, and possibly \pb). At the moment of this writing one might also speculate that all SLRs, except \al, \ca, \mn, and \fe, might be produced in a burst of nucleosynthesis during a NSM event, which would avoid the excesses on stable isotopes related to CCSNe. The status of our knowledge of these elusive but important phenomena is however still in its infancy and any further guess seems now to be by far premature.

\vspace{0.2cm}

Acknowledgments. 
We all are in debt with a very competent referee, willing to provide us 
with important advise on various aspects of the work, in a very constructive and helpful
way. DV is grateful to GSSI and to the INFN Section of Perugia for establishing the collaboration that permitted this work. MB, SP and OT are also grateful to the Department of Physics and Geology of Perugia for the financial support. SP thanks the Fondazione Cassa di Risparmio di Perugia for the three-year grant during which this work was performed.

\end{document}